\documentclass[12pt]{article}
\usepackage{graphicx}

\newcommand\pubnumber{WSU--HEP--XXYY}
\newcommand\pubdate{\today}

\def\Title#1{\begin{center} {\Large #1 } \end{center}}
\def\Author#1{\begin{center}{ \sc #1} \end{center}}
\def\Address#1{\begin{center}{ \it #1} \end{center}}

\newcommand\pubblock{\rightline{\begin{tabular}{l} \pubnumber\\
         \pubdate  \end{tabular}}}
\newenvironment{Abstract}{\begin{quotation}  }{\end{quotation}}
\newenvironment{Presented}{\begin{quotation} \begin{center}
             PRESENTED AT\end{center}\bigskip
      \begin{center}\begin{large}}{\end{large}\end{center} \end{quotation}}
\def\Acknowledgements{\bigskip  \bigskip \begin{center} \begin{large}
             \bf ACKNOWLEDGEMENTS \end{large}\end{center}}

\def\beq{\begin{equation}}
\def\eeq#1{\label{#1}\end{equation}}
\def\eeqn{\end{equation}}

\def\beqa{\begin{eqnarray}}
\def\eeqa#1{\label{#1}\end{eqnarray}}
\def\eeqan{\end{eqnarray}}

\let\bar=\overbar

\def\Dslash{\not{\hbox{\kern-4pt $D$}}}
\def\dslash{\not{\hbox{\kern-2pt $\del$}}}

\def\msb{{\bar{\ssstyle M \kern -1pt S}}}

\begin{document}
\begin{titlepage}
\pubblock

\vfill
\Title{$D$ leptonic and semileptonic decays}
\vfill
\Author{Hailong Ma (For BESIII Collaboration)
}
\Address{Institute of High Energy Physics, Chinese
        Academy of Sciences}
\vfill
\begin{Abstract}
Based on 2.92 fb$^{-1}$ data taken at the center-of-mass
energy $\sqrt s=3.773$ GeV with the BESIII detector,
we report recent results on
the decay constant $f_{D^+}$,
the hadronic form factors,
as well as the quark mixing matrix elements $|V_{cs(d)}|$,
which are extracted from analyses of the
leptonic decay $D^+ \to \mu^+\nu_\mu$ and the semileptonic
decays $D^0\to K(\pi)^-e^+\nu_e$,
$D^+\to K^0_L e^+\nu_e$,
$D^+\to K^-\pi^+e^+\nu_e$ and
$D^+\to \omega(\phi)e^+\nu_e$ at BESIII.
\end{Abstract}

\vfill
\begin{Presented}
The 7th International Workshop on Charm Physics (CHARM 2015)\\
Detroit, MI, 18-22 May, 2015
\end{Presented}
\vfill
\end{titlepage}
\def\thefootnote{\fnsymbol{footnote}}
\setcounter{footnote}{0}
%

\section{Introduction}

In the Standard Model, the $D^+$ mesons decay
into $\ell\nu_\ell$ via a virtual $W^+$ boson. The decay rate of
the leptonic decays
$D^+\to \ell^+\nu_\ell$ can be parameterized by the $D^+$ decay
constant $f_{D^+}$ via
\begin{equation}
\Gamma(D^+\to\ell^+\nu_\ell)=\frac{G_F^2}{8\pi}|V_{cd}|^2
f_{D^+}
m_\ell^2 m_{D^+} (1-\frac{m_\ell^2}{m_{D^+}^2}),
\end{equation}
where $G_F$ is the Fermi coupling constant, $|V_{cd}|$ is the
quark mixing matrix element between the two quarks $c\bar d$, $m_\ell$ and
$m_{D^+}$ are the lepton and $D^+$ masses.

On the other hand, the $D$ semileptonic decays can be
parameterized by the quark mixing matrix element and the form factor
of hadronic weak current simply, thus providing an ideal window to probe for
the weak and strong effects.
For example, the differential decay rates of $D\to K(\pi)e^+\nu_e$
can be simply written as
\begin{equation}
\frac{d\Gamma}{dq^2} =\frac{G_F^2}{24\pi^3}|V_{cs(d)}|^2
p_{K(\pi)}^3 |f_{+}^{K(\pi)}(q^2)|^2,
\end{equation}
where $G_F$ is the Fermi coupling constant, $|V_{cs(d)}|$ is the
quark mixing matrix element between the two quarks $c\bar s(\bar d)$,
$p_{K(\pi)}$ is the kaon(pion) momentum in the $D^0$ rest frame,
$f_{+}^{K(\pi)}(q^2)$ is the form factor of hadronic weak
current depending on the square of the four momentum transfer
$q=p_D-p_{K(\pi)}$.

In 2010 and 2011, BESIII \cite{bes3} accumulated 2.92 fb$^{-1}$ data at $\sqrt s$ = 3.773
GeV \cite{lum}, where $e^+e^-\to \psi(3770)\to$ $D^0\bar D^0$ or $D^+D^-$
is produced predominantly.
Based on the studies of the leptonic and semileptonic decays of $D^0$ and $D^+$ mesons,
the $D^+$ decay constant, the hadronic form factors
or the quark mixing matrix elements $|V_{cd(s)}|$ can be extracted accurately.
These will validate the LQCD calculations of
the $D^+$ decay constant and the hadronic form factors
or test the unitarity of the quark mixing matrix at higher accuracies.
They are also helpful to improve the measurement
precisions in the experimental studies of the leptonic and semileptonic decays
of $B$ mesons indirectly.
Herein, we report recent results on the studies of the
leptonic decay $D^+ \to \mu^+\nu_\mu$ and the semileptonic
decays $D^0\to K(\pi)^-e^+\nu_e$,
$D^+\to K^0_L e^+\nu_e$,
$D^+\to K^-\pi^+e^+\nu_e$ and
$D^+\to \omega(\phi)e^+\nu_e$ at BESIII.
Throughout the proceeding, charge conjugate is implied.


\section{Leptonic decay \cite{bes3_rg}}

\begin{figure*}[htbp]
\begin{center}
\includegraphics[width=0.4\textwidth,height=0.3\textwidth]{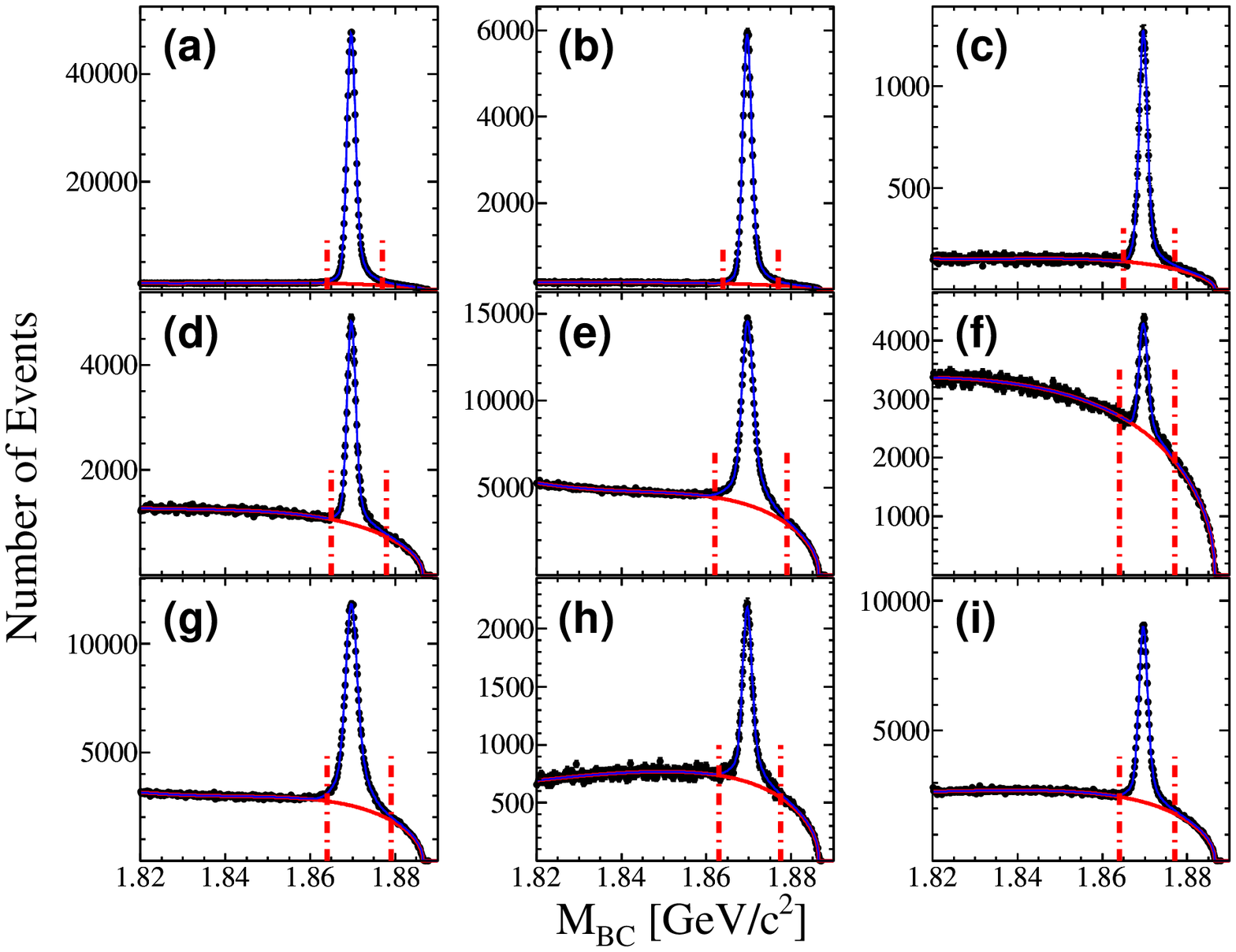}
\hspace{2.0cm}
\includegraphics[width=0.4\textwidth,height=0.3\textwidth]{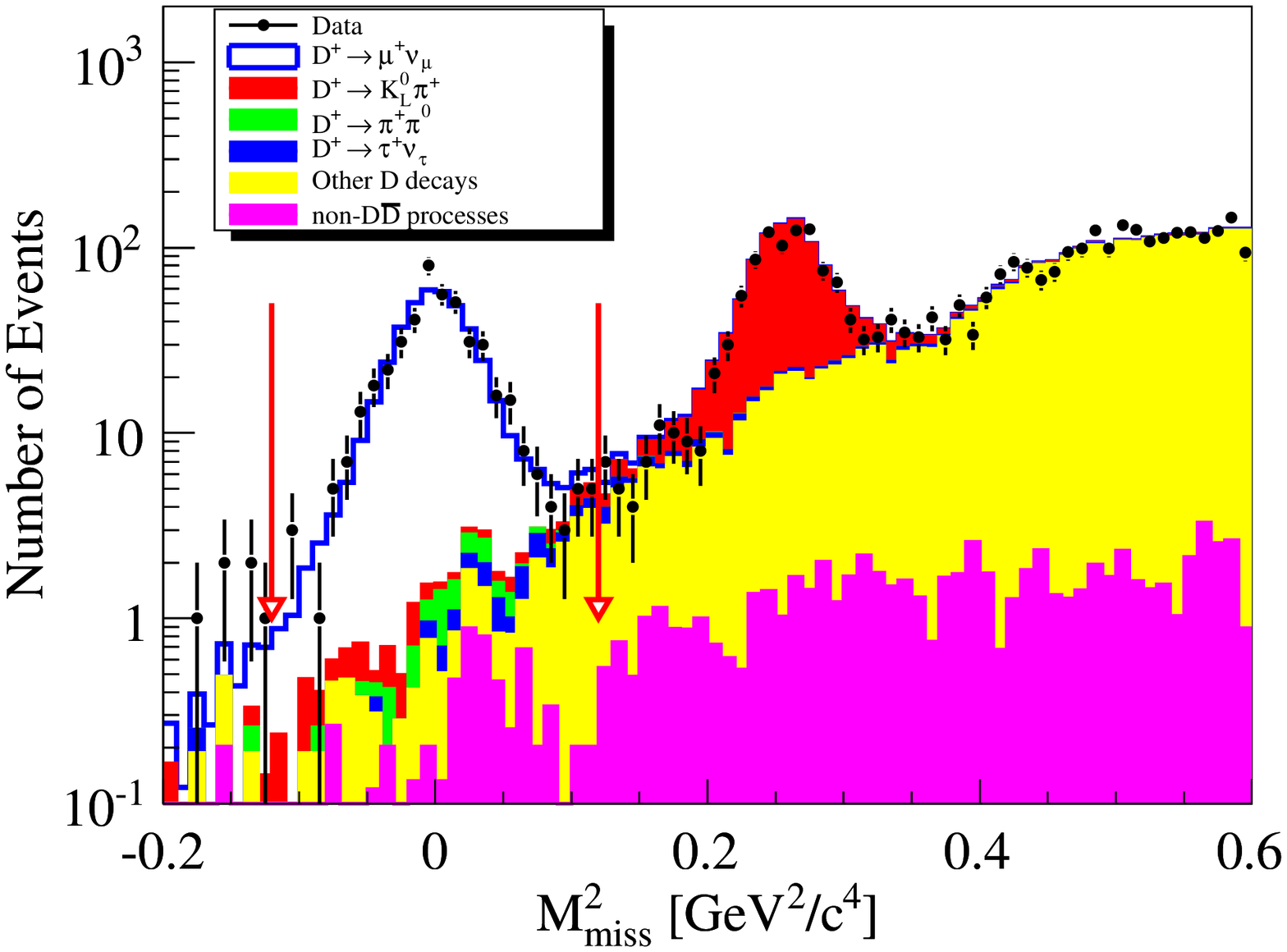}
\caption{ (left  side) Fits to the $M_{\rm BC}$
spectra for the singly tagged $D^-$ candidates
(the signal region is marked by the pair of arrows in each sub-figure).
(right side) $M^2_{\rm
miss}$ distribution for $D^+ \to \mu^+\nu_\mu$ candidates.}
\label{fig:1}
\end{center}
\end{figure*}

To investigate the leptonic decay $D^+ \to \mu^+\nu_\mu$,
we reconstruct the singly tagged $D^-$ mesons using 9 hadronic decays.
Figure \ref{fig:1} (left side) shows the fits to the
beam-energy-constrained mass ($M_{\rm BC}$) spectra of the
(a) $K^+\pi^-\pi^-$, (b) $K_S^0\pi^-$,
(c) $K_S^0K^-$, (d) $K^+K^-\pi^-$, (e) $K^+\pi^-\pi^-\pi^0$, (f)
$\pi^-\pi^-\pi^+$, (g) $K_S^0\pi^-\pi^0$, (h)
$K^+\pi^-\pi^-\pi^-\pi^+$ and (i) $K_S^0\pi^-\pi^-\pi^+$
combinations, which yield $(170.31\pm0.34)\times 10^4$
singly tagged $D^-$ mesons.

Figure \ref{fig:1} (right side) shows the $M^2_{\rm miss}$
distribution of the candidates for $D^+ \to \mu^+\nu_\mu$,
which are selected in the systems against the
singly tagged $D^-$ mesons.
We obtain $409\pm21$ signals of $D^+\to\mu^+\nu_\mu$,
which yields
the branching fraction
$$B(D^+ \to \mu^+\nu_\mu)=(3.71\pm0.19_{\rm stat.}\pm0.06_{\rm sys.})\times10^{-4}.$$

\begin{figure*}[htbp]
\begin{center}
\includegraphics[width=0.4\textwidth,height=0.3\textwidth]{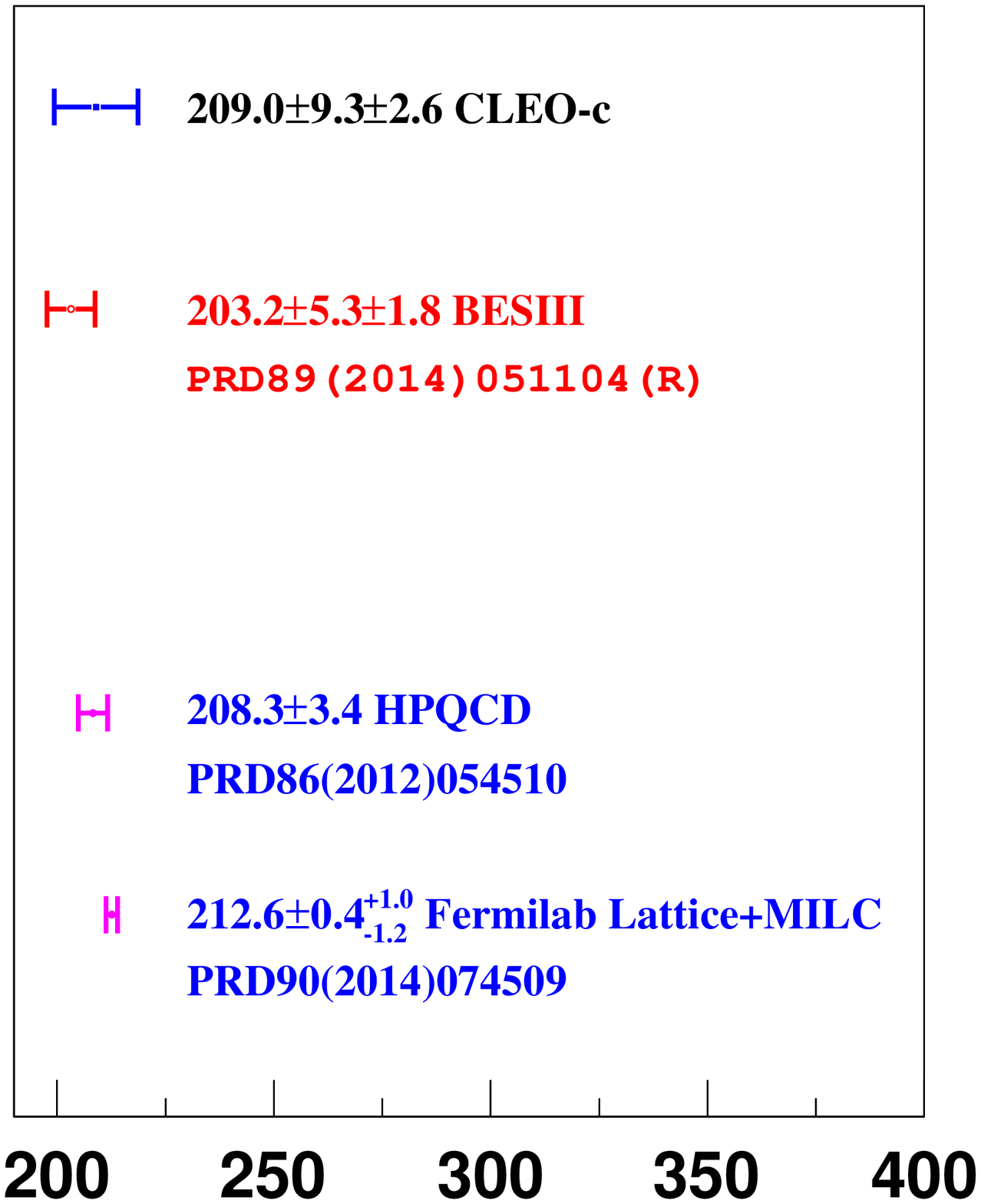}
\put(-110.0,0){\bf $f_{D^+}$ (MeV)}
\caption{Comparison of the $D^+$ decay constant.}
\label{fig:3}
\end{center}
\end{figure*}

Using the measured $B(D^+ \to \mu^+\nu_\mu)$ and the quark mixing matrix
element $|V_{cd}|$ from a global Standard Model fit \cite{pdg2014},
we determine the $D^+$ decay constant
$$f_{D^+}=203.2\pm5.3_{\rm stat.}\pm1.8_{\rm sys.}~\rm MeV.$$
The $B(D^+ \to \mu^+\nu_\mu)$ and $f_{D^+}$ measured at BESIII
are consistent within errors with those measured at BESI
\cite{bes1_muv}, BESII \cite{bes2_muv} and CLEO-c \cite{cleo_muv}, but with the best precision.
Figure
\ref{fig:3} compares the $f_{D^+}$ measured at BESIII and CLEO-c
as well as those calculated by recent theories.

So far, the quark mixing matrix element $|V_{cd}|$ has been measured
though experimental studies of the semileptonic decay $D\to \pi
\ell^+\nu_\ell$ or measurement of charm production cross section of
$\nu \bar \nu$ interaction, among which the best measurement
precision is 4.8\% \cite{pdg2014}. By using the measured $B(D^+ \to
\mu^+\nu_\mu)$ and the Lattice QCD calculation on $f_{D^+}$
\cite{pqcd}, we determine
$$|V_{cd}|=0.2210\pm0.058_{\rm stat.}\pm0.047_{\rm sys.},$$
which has the best precision in the world to date.

\section{Semileptonic decays}

\subsection{$D^0$ semileptonic decays \cite{bes3_kpiev}}

\begin{figure*}[htbp]
\begin{center}
\includegraphics[width=0.4\textwidth,height=0.3\textwidth]{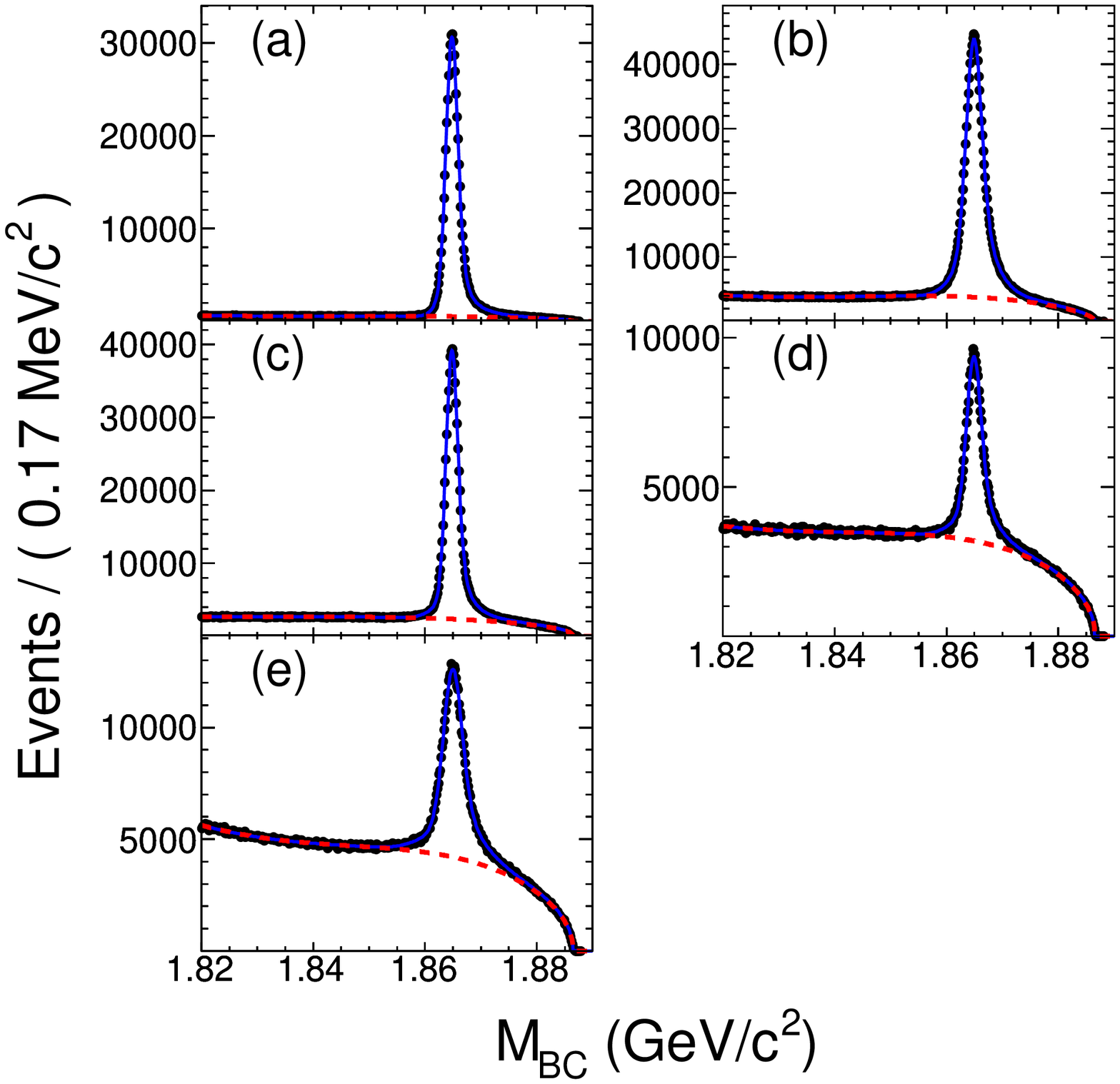}
\hspace{2.0cm}
\includegraphics[width=0.4\textwidth,height=0.3\textwidth]{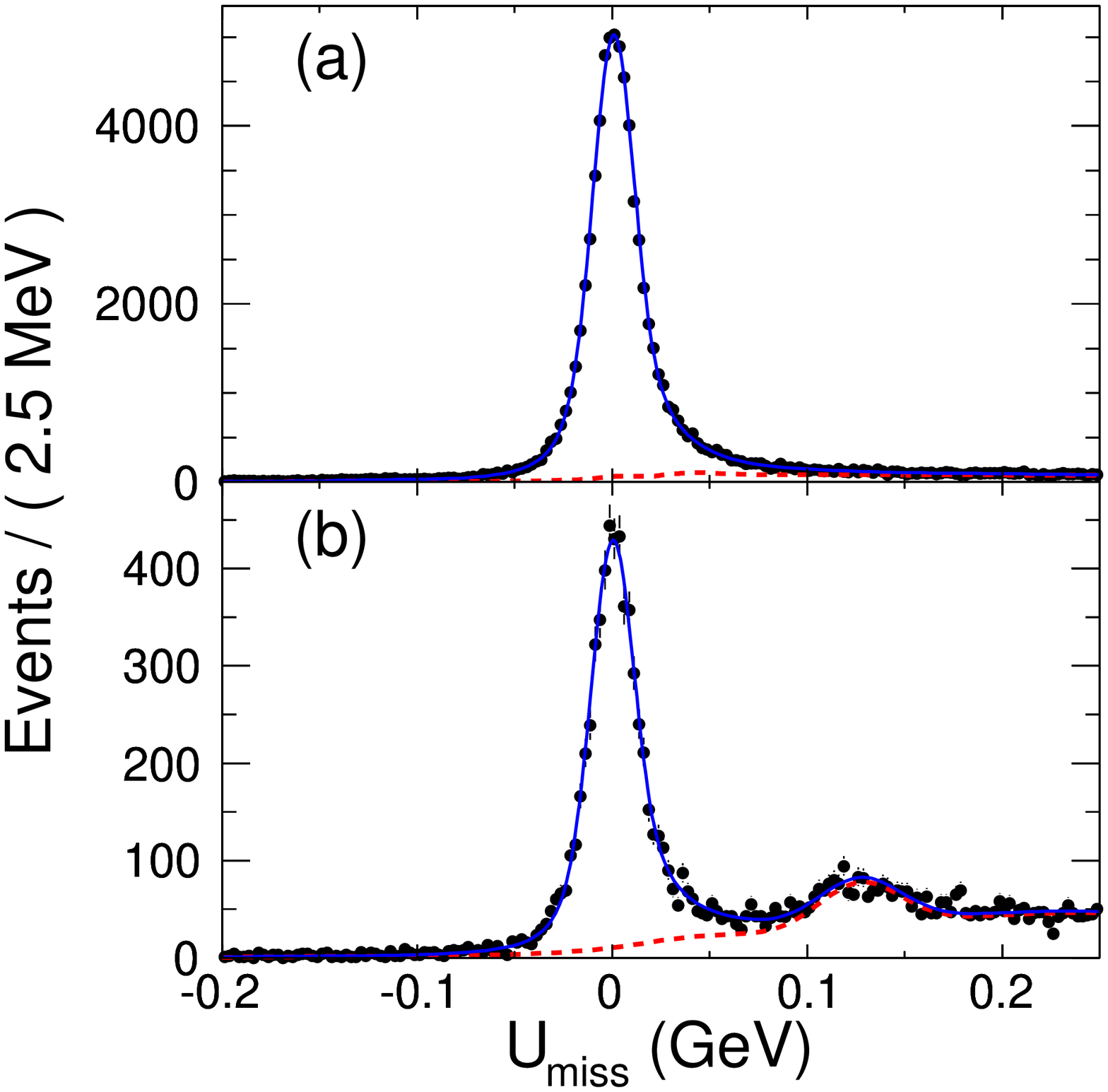}
\caption{ (left  side) Fits to the $M_{\rm BC}$
spectra for the singly tagged $\bar D^0$ candidates.
(right side) Fits to the $U_{\rm
miss}$ distributions for (a) $D^0\to K^-e^+\nu_e$ and (b) $D^0\to \pi^-e^+\nu_e$
candidates.} \label{fig:4}
\end{center}
\end{figure*}

To investigate the semileptonic decays $D^0\to K(\pi)^-e^+\nu_e$,
we reconstruct the singly tagged $\bar D^0$ mesons
using 5 hadronic decays.
Figure \ref{fig:4} (left side) shows the fits to the $M_{\rm BC}$ spectra of the
(a) $K^+\pi^-$,
(b) $K^+\pi^-\pi^0$,
(c) $K^+\pi^-\pi^-\pi^+$,
(d) $K^+\pi^-\pi^-\pi^+\pi^0$ and
(e) $K^+\pi^-\pi^0\pi^0$ combinations.
$(279.33\pm0.37)\times 10^4$ singly tagged $\bar D^0$ mesons are accumulated.

Figure \ref{fig:4} (right side) shows the fits to the $U_{\rm
miss}$ distributions of the candidates for $D^0\to K^-e^+\nu_e$ and $D^0\to
\pi^-e^+\nu_e$,
which are selected in the systems against the
singly tagged $D^-$ mesons.
From the fits, we obtain $70727\pm278$ and
$6297\pm87$ signals of $D^0\to K^-e^+\nu_e$ and
$D^0\to \pi^-e^+\nu_e$. Based on these, we determine the branching
fractions
$$B(D^0\to K^-e^+\nu_e)=
(3.505\pm0.014_{\rm stat.}\pm0.033_{\rm sys.})\%$$ and
$$B(D^0\to \pi^-e^+\nu_e)=
(0.2950\pm0.0041_{\rm stat.}\pm0.0026_{\rm sys.})\%,$$ respectively.
The $B(D^0\to K^-e^+\nu_e)$ and $B(D^0\to \pi^-e^+\nu_e)$
measured at BESIII are consistent within errors with those measured at
BESII \cite{bes2_kpiev}, CLEO-c \cite{cleo_kpiev}, BELLE
\cite{belle_kpiev} and BABAR \cite{babar_kev,babar_piev},
but with the best precision.

\begin{figure*}[htbp]
\begin{center}
\includegraphics[width=0.4\textwidth,height=0.2\textwidth]{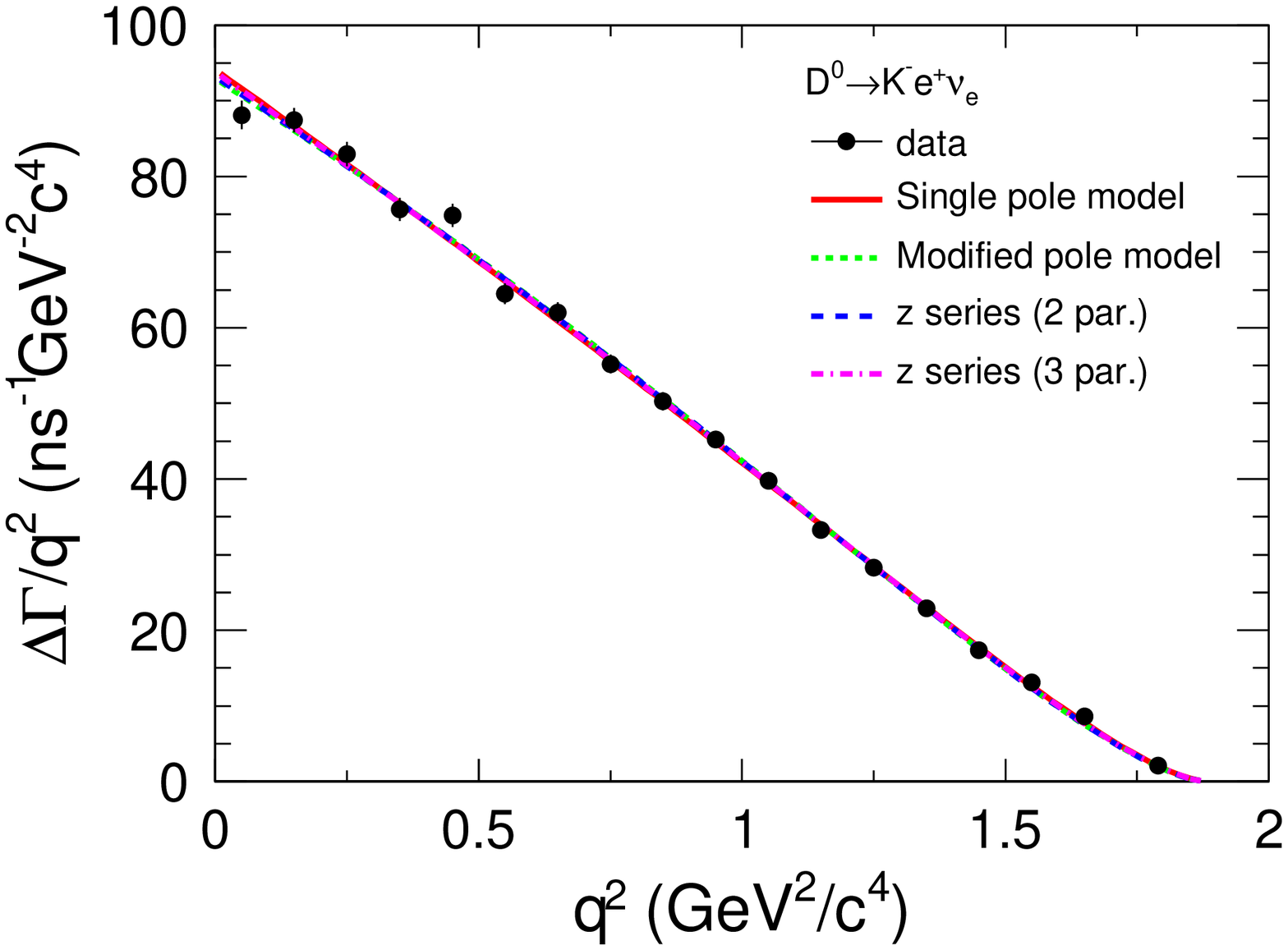}
\hspace{2.0cm}
\includegraphics[width=0.4\textwidth,height=0.2\textwidth]{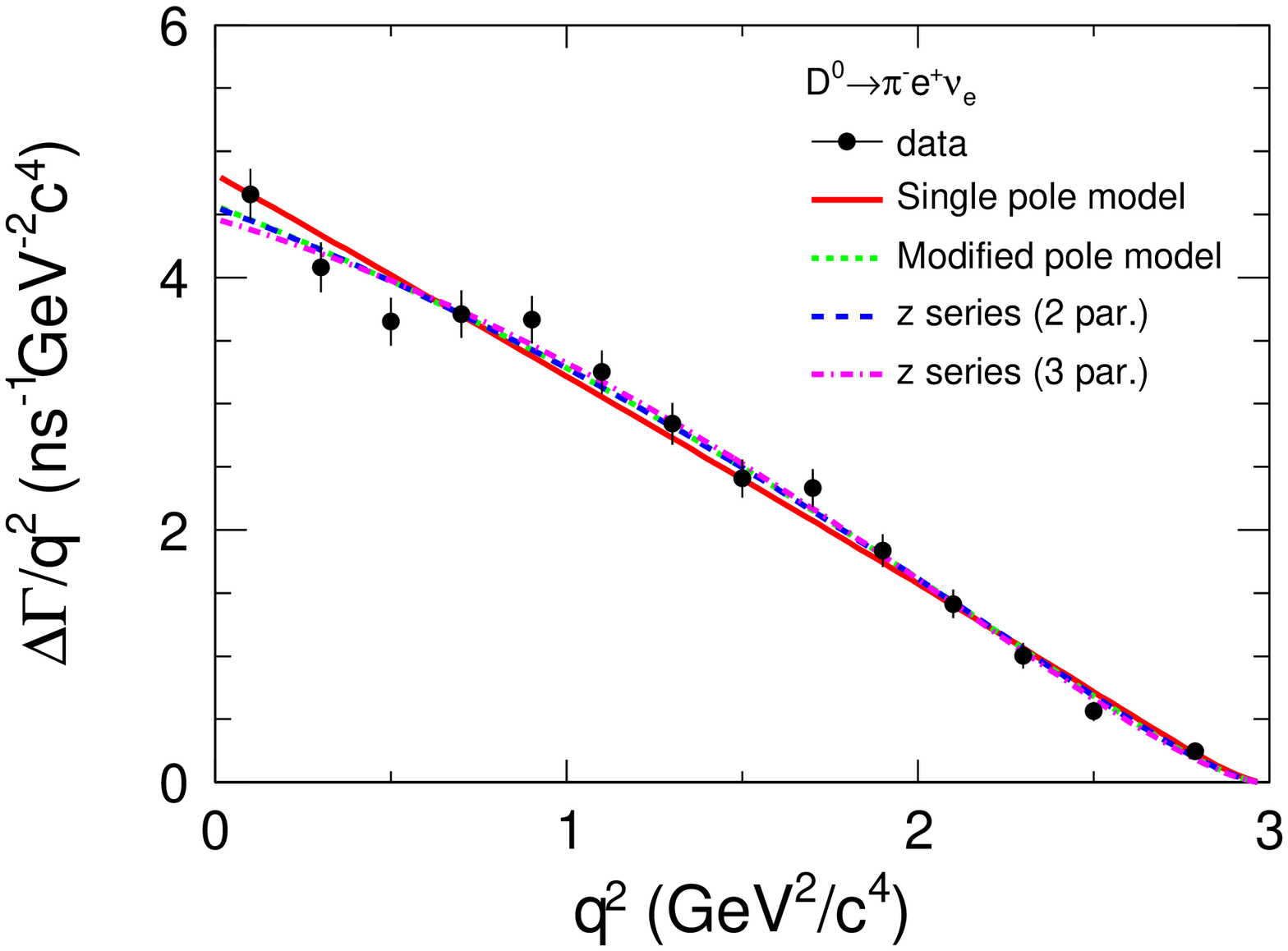}
\includegraphics[width=0.4\textwidth,height=0.2\textwidth]{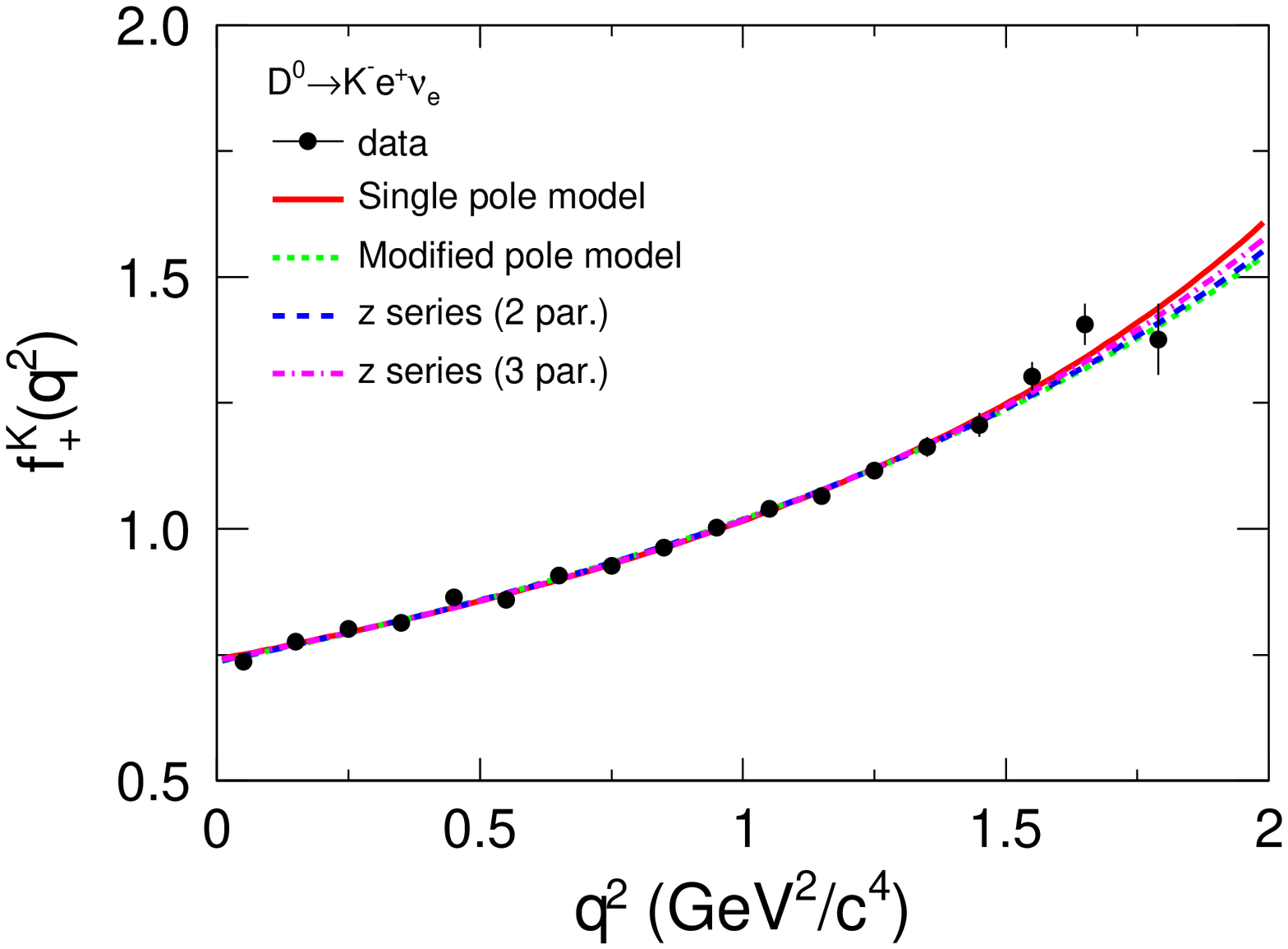}
\hspace{2.0cm}
\includegraphics[width=0.4\textwidth,height=0.2\textwidth]{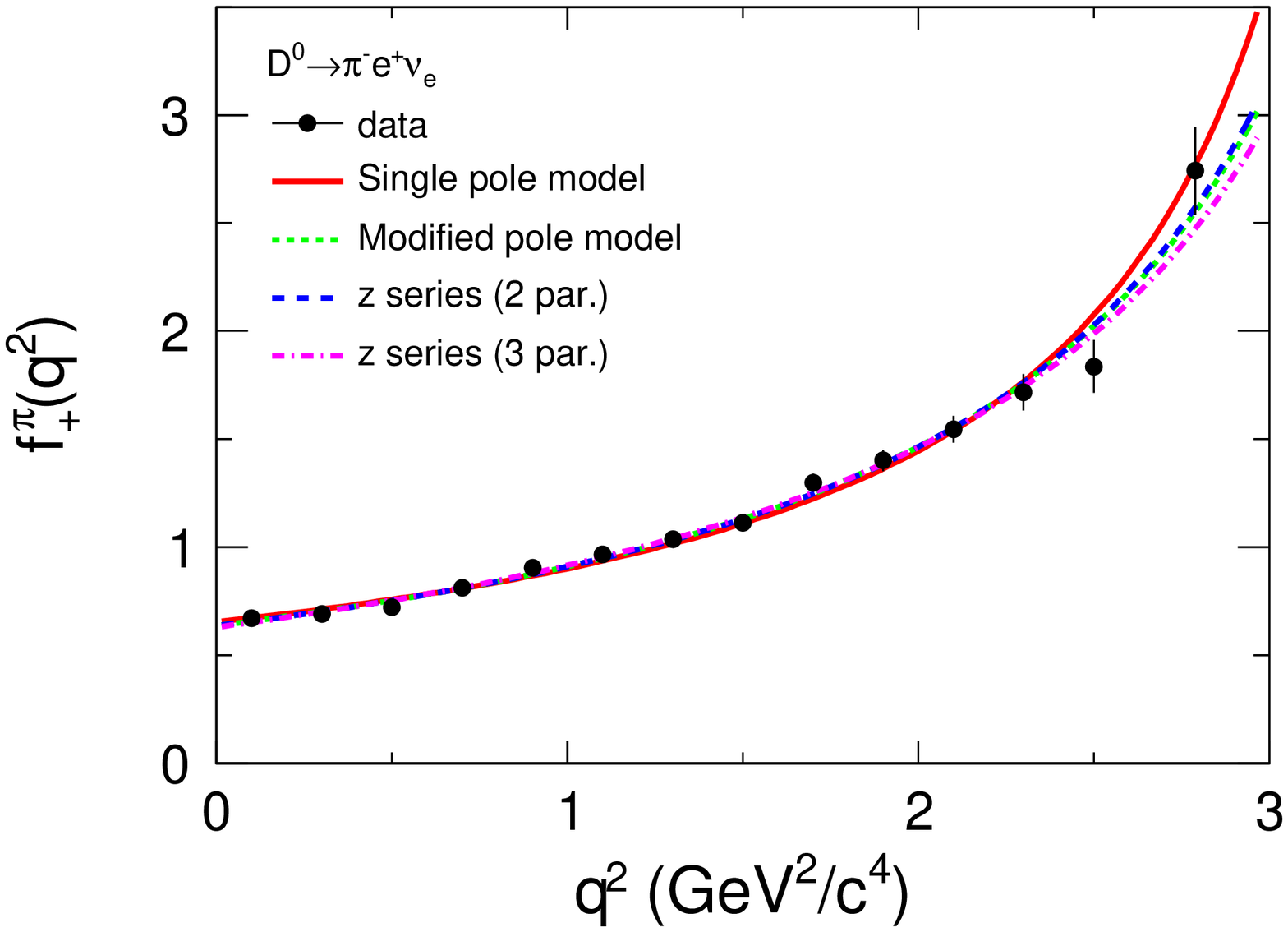}
\caption{The fits to the partial widths and the projections on
the form factors of $D^0\to K^-e^+\nu_e$ (left side) and
$D^0\to \pi^-e^+\nu_e$ (right side).}
\label{fig:5}
\end{center}
\end{figure*}

Figure \ref{fig:5} shows the
fits to the partial widths and the projections on
the form factors of $D^0\to K^-e^+\nu_e$ and
$D^0\to \pi^-e^+\nu_e$
using the Simple Pole model \cite{simple},
the Modified Pole model \cite{simple},
the ISGW2 model \cite{isgw2},
the two-parameter series expansion (Series.2.Par.) \cite{expansion} and
the three-parameter series expansion (Series.3.Par.) \cite{expansion}.
From the fits, we obtain the extracted parameters
of different models, which are summarized in Table \ref{tab:1}.

\begin{table*}[hbtp]
\scriptsize
\begin{center}
\caption{\small
Summary of the extracted parameters from the fits to the partial widths,
where the first errors are statistical and the second systematic.}
\begin{tabular}{|l|c|c|c|c|} \hline
Model
& \multicolumn{2}{|c|} {$D^0\to K^-e^+\nu_e$}
& \multicolumn{2}{|c|} {$D^0\to \pi^-e^+\nu_e$} \\ \hline
Simple Pole &
$f_+^K(0)|V_{cs}|$&$0.7209\pm0.0022\pm0.0033$&
$f_+^\pi(0)|V_{cd}|$&$0.1475\pm0.0014\pm0.0005$\\
&
$M_{\rm pole}$&$1.9207\pm0.0103\pm0.0069$&
$M_{\rm pole}$&$1.9114\pm0.0118\pm0.0038$ \\ \hline
Modified Pole &
$f_+^K(0)|V_{cs}|$&$0.7163\pm0.0024\pm0.0034$&
$f_+^\pi(0)|V_{cd}|$&$0.1437\pm0.0017\pm0.0008$\\
&
$\alpha$&$0.3088\pm0.0195\pm0.0129$&
$\alpha$&$0.2794\pm0.0345\pm0.0113$ \\ \hline
ISGW2 &
$f_+^K(0)|V_{cs}|$&$0.7139\pm0.0023\pm0.0034$&
$f_+^\pi(0)|V_{cd}|$&$0.1415\pm0.0016\pm0.0006$\\
&
$r_{\rm ISGW2}$&$1.6000\pm0.0141\pm0.0091$&
$r_{\rm ISGW2}$&$2.0688\pm0.0394\pm0.0124$ \\ \hline
Series.2.Par. &
$f_+^K(0)|V_{cs}|$&$0.7172\pm0.0025\pm0.0035$&
$f_+^\pi(0)|V_{cd}|$&$0.1435\pm0.0018\pm0.0009$\\
&
$r_1$&$-2.2278\pm0.0864\pm0.0575$&
$r_1$&$-2.0365\pm0.0807\pm0.0260$ \\ \hline
Series.3.Par. &
$f_+^K(0)|V_{cs}|$&$0.7196\pm0.0035\pm0.0041$&
$f_+^\pi(0)|V_{cd}|$&$0.1420\pm0.0024\pm0.0010$\\
&
$r_1$&$-2.3331\pm0.1587\pm0.0804$&
$r_1$&$-1.8434\pm0.2212\pm0.0690$ \\
&
$r_2$&$ 3.4223\pm3.9090\pm2.4092$&
$r_2$&$-1.3871\pm1.4615\pm0.4677$ \\ \hline
\end{tabular}
\label{tab:1}
\end{center}
\end{table*}

\begin{figure*}[htbp]
\begin{center}
\includegraphics[width=0.4\textwidth,height=0.4\textwidth]{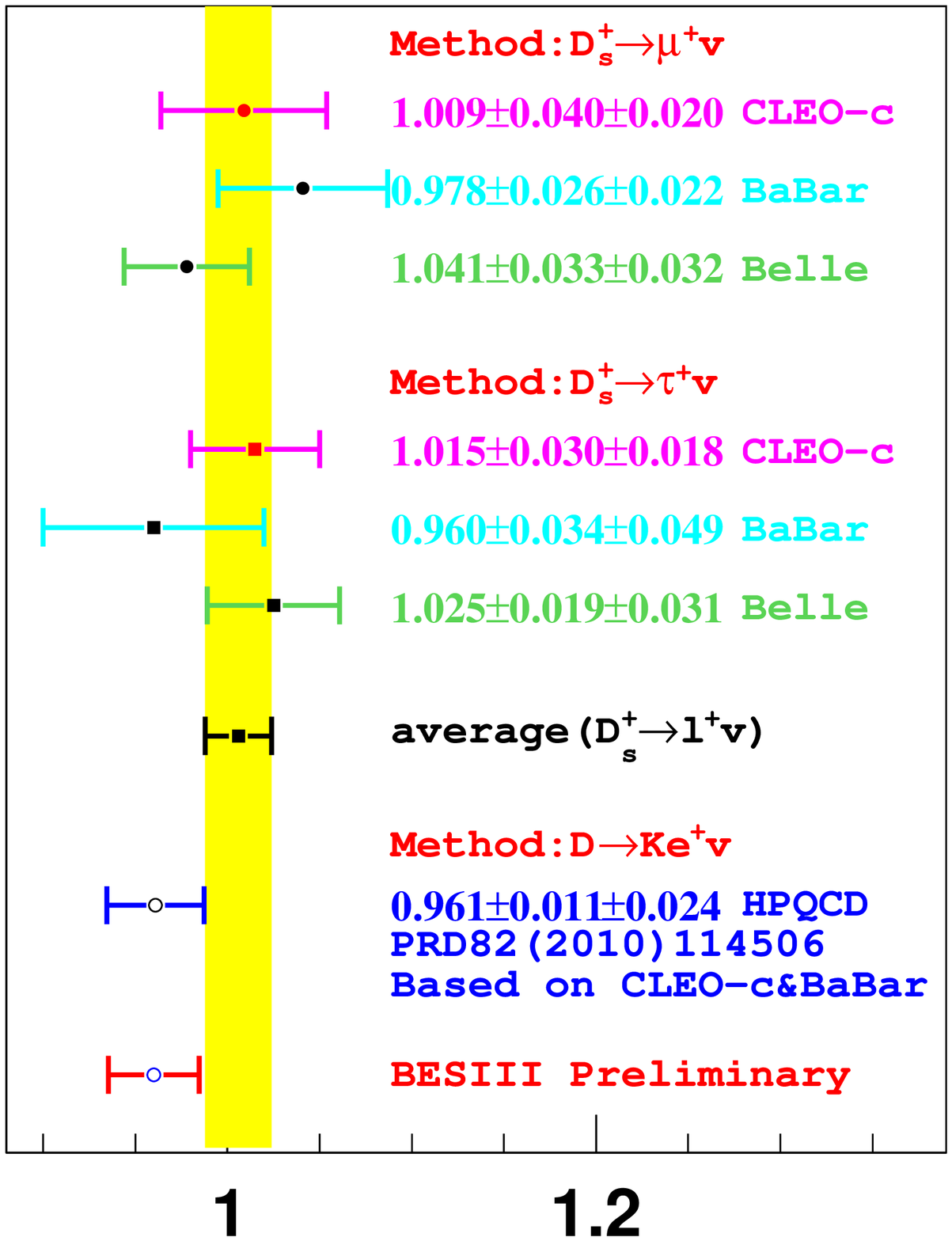}
\put(-100.0,8){\bf $|V_{cs}|$} \hspace{2.0cm}
\includegraphics[width=0.4\textwidth,height=0.4\textwidth]{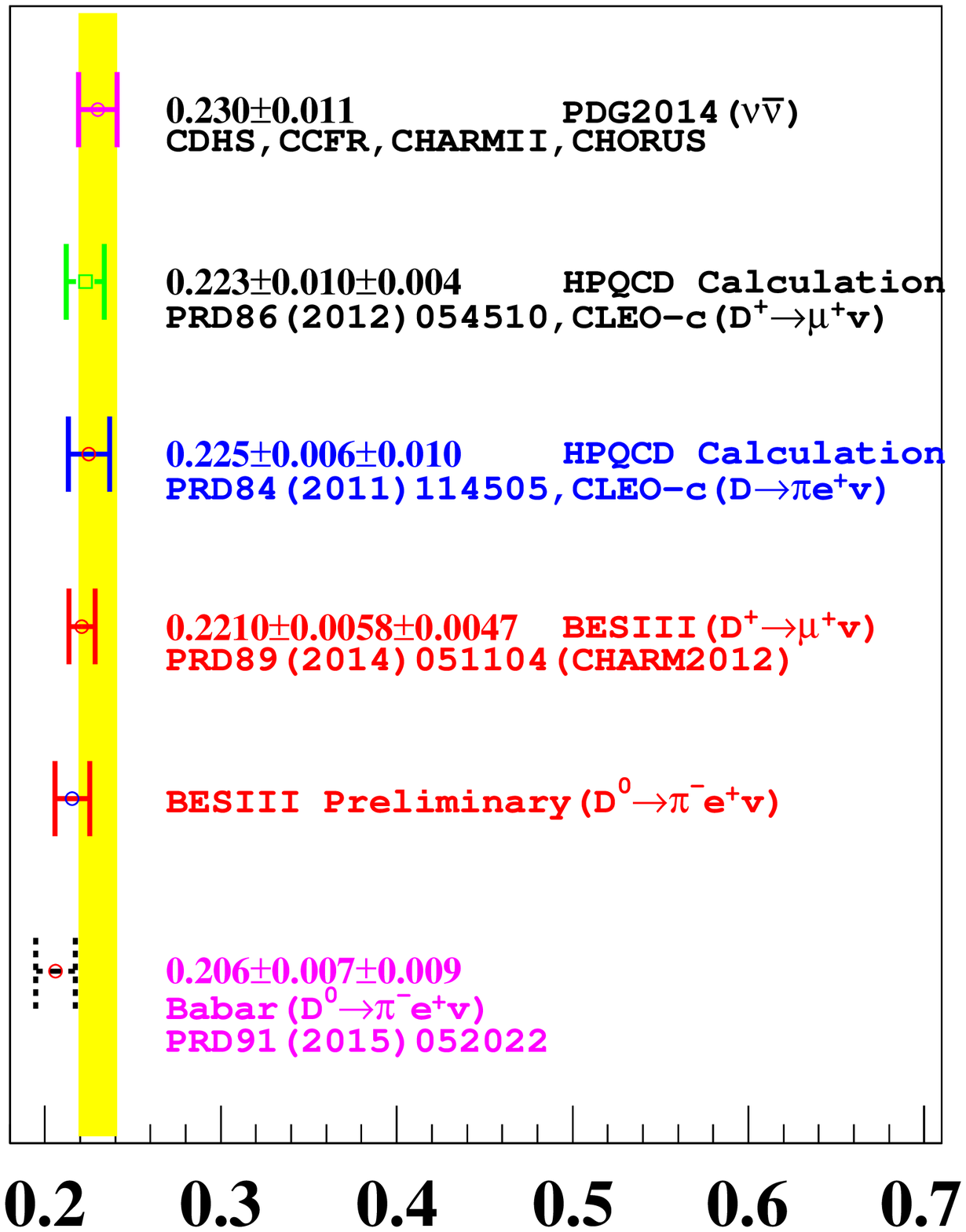}
\put(-100.0,8){\bf $|V_{cd}|$} \caption{Comparison
of the extracted $|V_{cs(d)}|$.} \label{fig:6}
\end{center}
\end{figure*}

With the extracted $f_+^{K(\pi)}(0)|V_{cs(d)}|$ and the expected $f_+^{K(\pi)}(0)$
by LQCD \cite{LQCD_ff}, we determine the quark mixing matrix element $|V_{cs(d)}|$.
Figure
\ref{fig:6} compares the $|V_{cs(d)}|$ extracted at BESIII
with the ones from other experiments.

\subsection{$D^+$ Semileptonic decays}

To study the semileptonic decays $D^+\to K^0_Le^+\nu_e$,
$D^+\to K^-\pi^+e^+\nu_e$ and $D^+\to \omega(\phi)e^+\nu_e$,
we reconstruct the singly tagged $D^-$ mesons using 6 hadronic
decays of
$K^+\pi^-\pi^-$, $K^+\pi^-\pi^-\pi^0$,
$K_S^0\pi^-$, $K_S^0\pi^-\pi^0$,
$K_S^0\pi^-\pi^-\pi^+$ and $K^+K^-\pi^-$.
About 1.6 millions of singly tagged $D^-$ mesons are accumulated \cite{bes3_omgev}.
Based on these, we study
the semileptonic decays $D^+\to K^0_Le^+\nu_e$,
$D^+\to K^-\pi^+e^+\nu_e$ and $D^+\to \omega(\phi)e^+\nu_e$.

\subsubsection{Analysis of $D^+\to K^0_Le^+\nu_e$}
\begin{figure*}[htbp]
\begin{center}
\includegraphics[width=0.25\textwidth,height=0.18\textwidth]{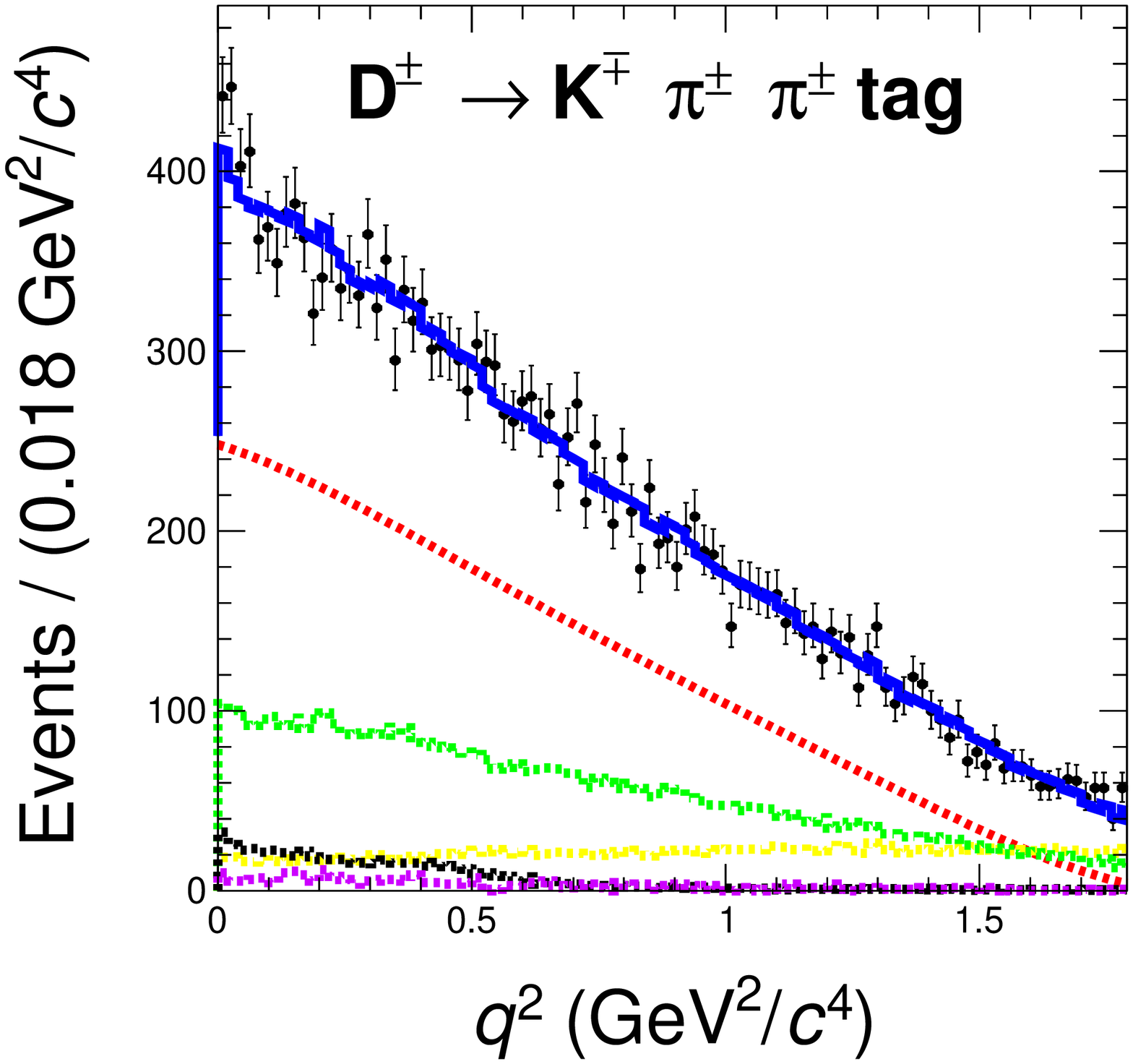}
\includegraphics[width=0.25\textwidth,height=0.18\textwidth]{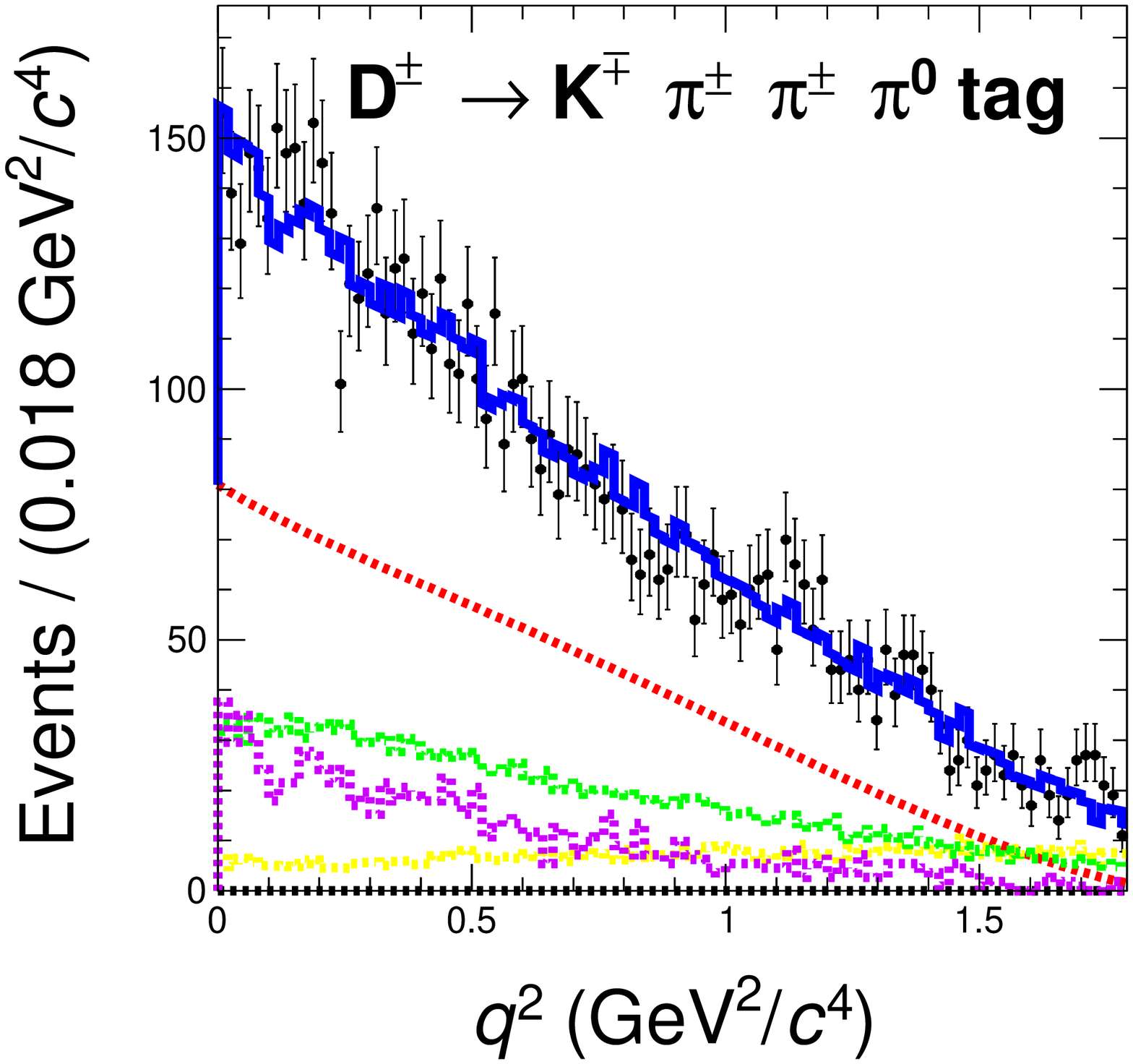}
\includegraphics[width=0.25\textwidth,height=0.18\textwidth]{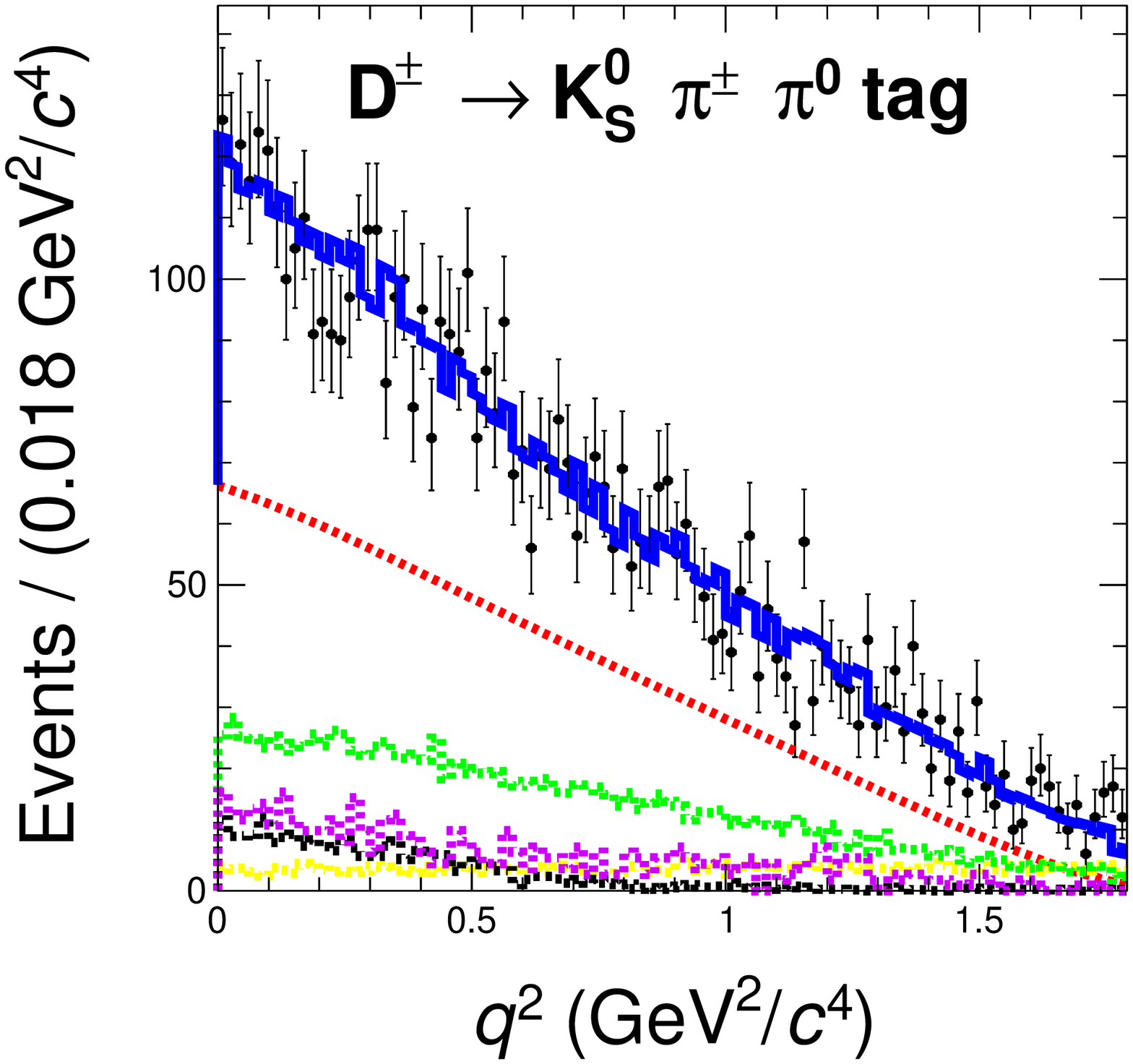}
\includegraphics[width=0.25\textwidth,height=0.18\textwidth]{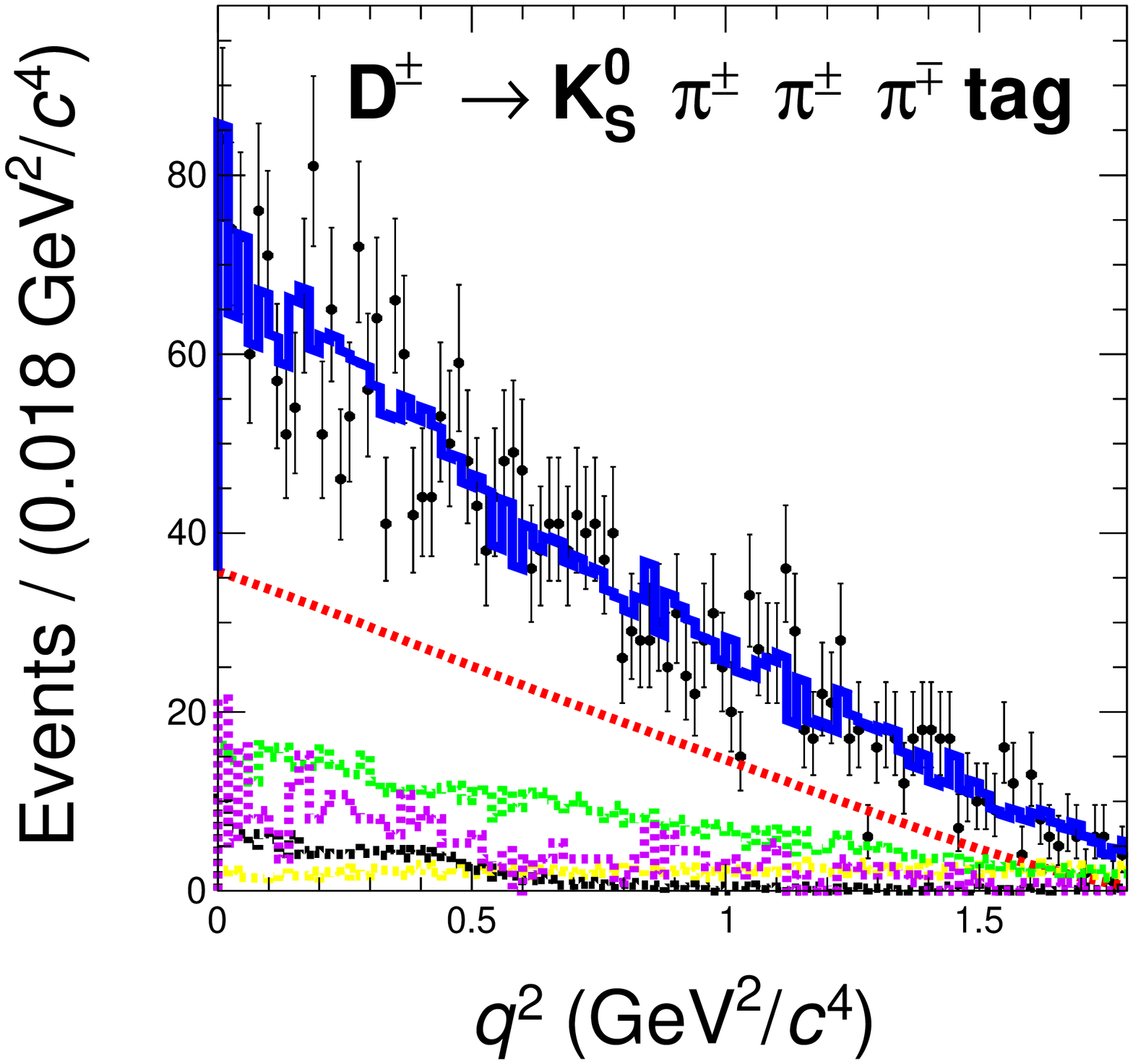}
\includegraphics[width=0.25\textwidth,height=0.18\textwidth]{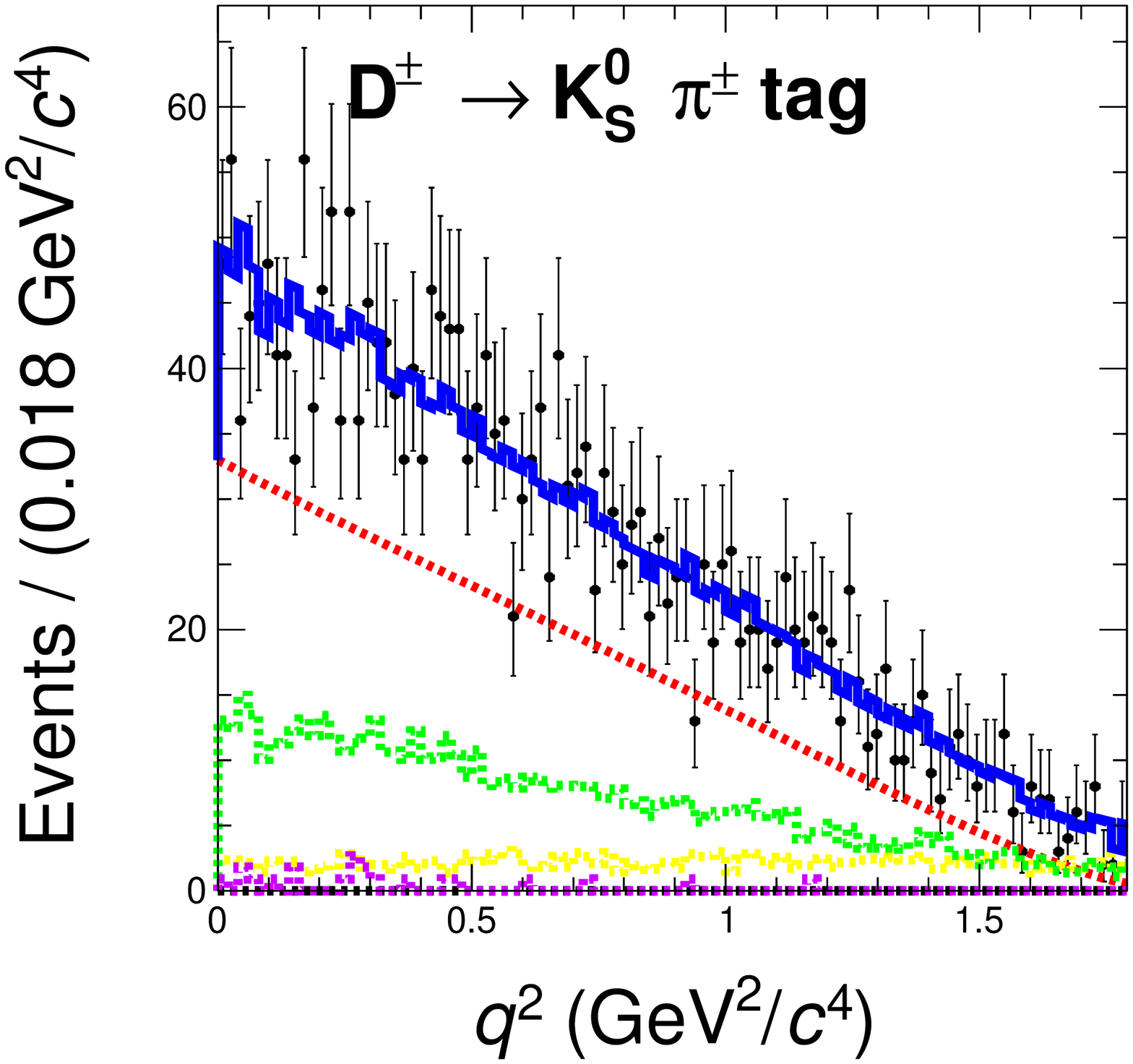}
\includegraphics[width=0.25\textwidth,height=0.18\textwidth]{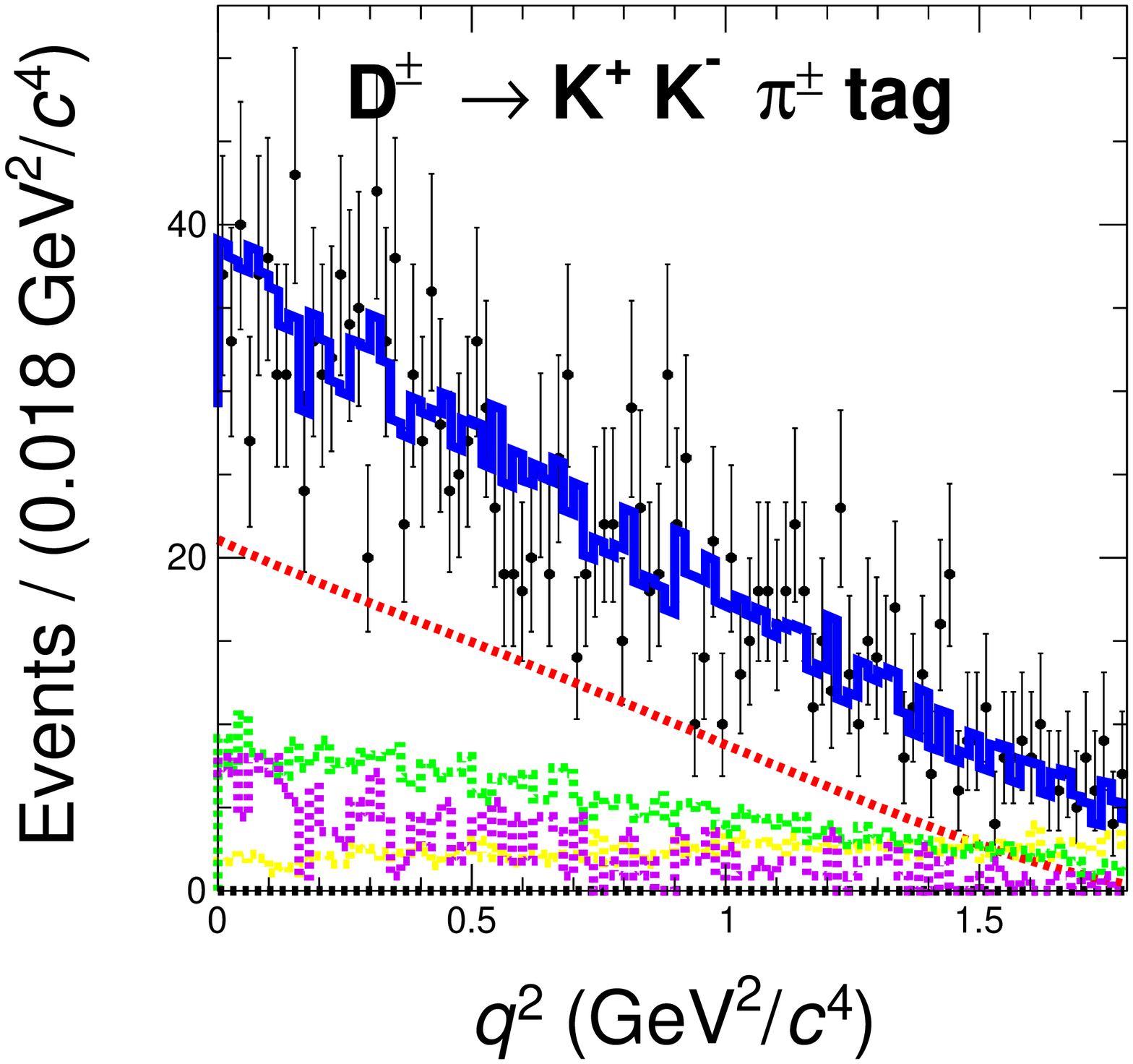}
\caption{Simultaneous fit to the event density $I(q^2)$ for different tag modes,
where the points with error bars are data and blue curves are the fits.
The violet, yellow, green and black curves refer to the different background sources.
}
\label{fig:klev}
\end{center}
\end{figure*}

Although $K^0_L$ flights long distance,
it interacts with the Electron Magnetic Cluster of BESIII and
deposits a portion of energy, thus leaving some position
information.
So, after reconstructing all other particles in the final states,
the $K^0_L$ mesons can be inferred with the position
information and constraining the $U_{\rm miss}$ of the candidates
to zero. We obtain about 24 thousands of signals of $D^+\to K^0_Le^+\nu_e$,
based on which we determine the branching fraction
$$B(D^+\to K^0_Le^+\nu_e)=(4.482\pm0.027_{\rm stat.}\pm0.103_{\rm sys.})\%$$
and the CP asymmetry
$$A_{\rm CP}^{D^+\to K^0_Le^+\nu_e}=(-0.59\pm0.60_{\rm stat.}\pm1.50_{\rm sys.})\%,$$
supporting that there is no CP asymmetry in this decay.
In addition, simultaneous fit to the event density
$I(q^2)$ for different tag modes with the two-parameter series expansion
is performed, as shown in Fig. \ref{fig:klev},
which yields the product of $f_+^{K}(0)|V_{cs}|=0.728\pm0.006_{\rm stat.}\pm0.011_{\rm sys.}$.
These are made for the first time.

\subsubsection{$D^+\to K^-\pi^+e^+\nu_e$}

\begin{figure*}[htbp]
\begin{center}
\includegraphics[width=0.25\textwidth,height=0.18\textwidth]{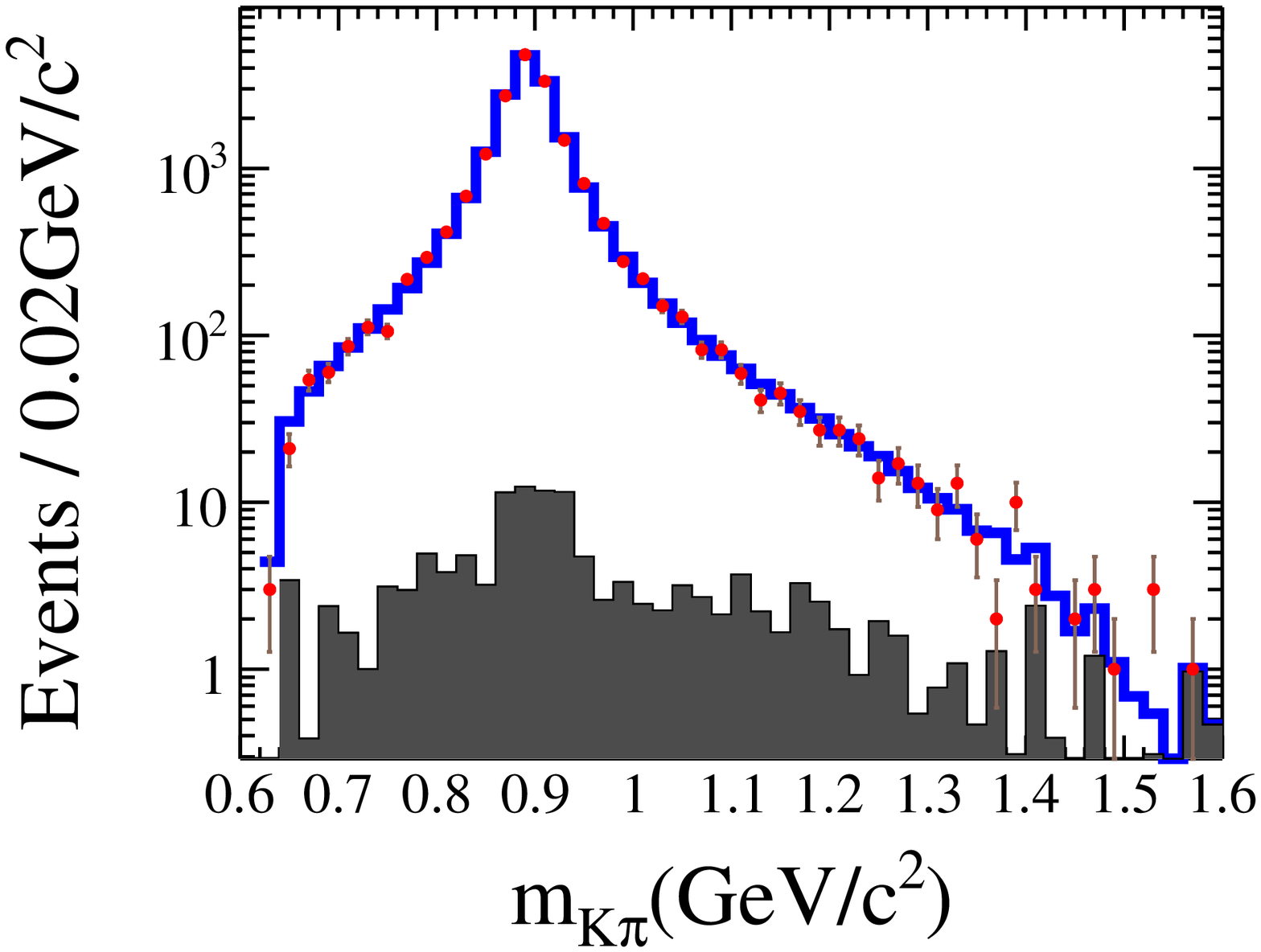}
\includegraphics[width=0.25\textwidth,height=0.18\textwidth]{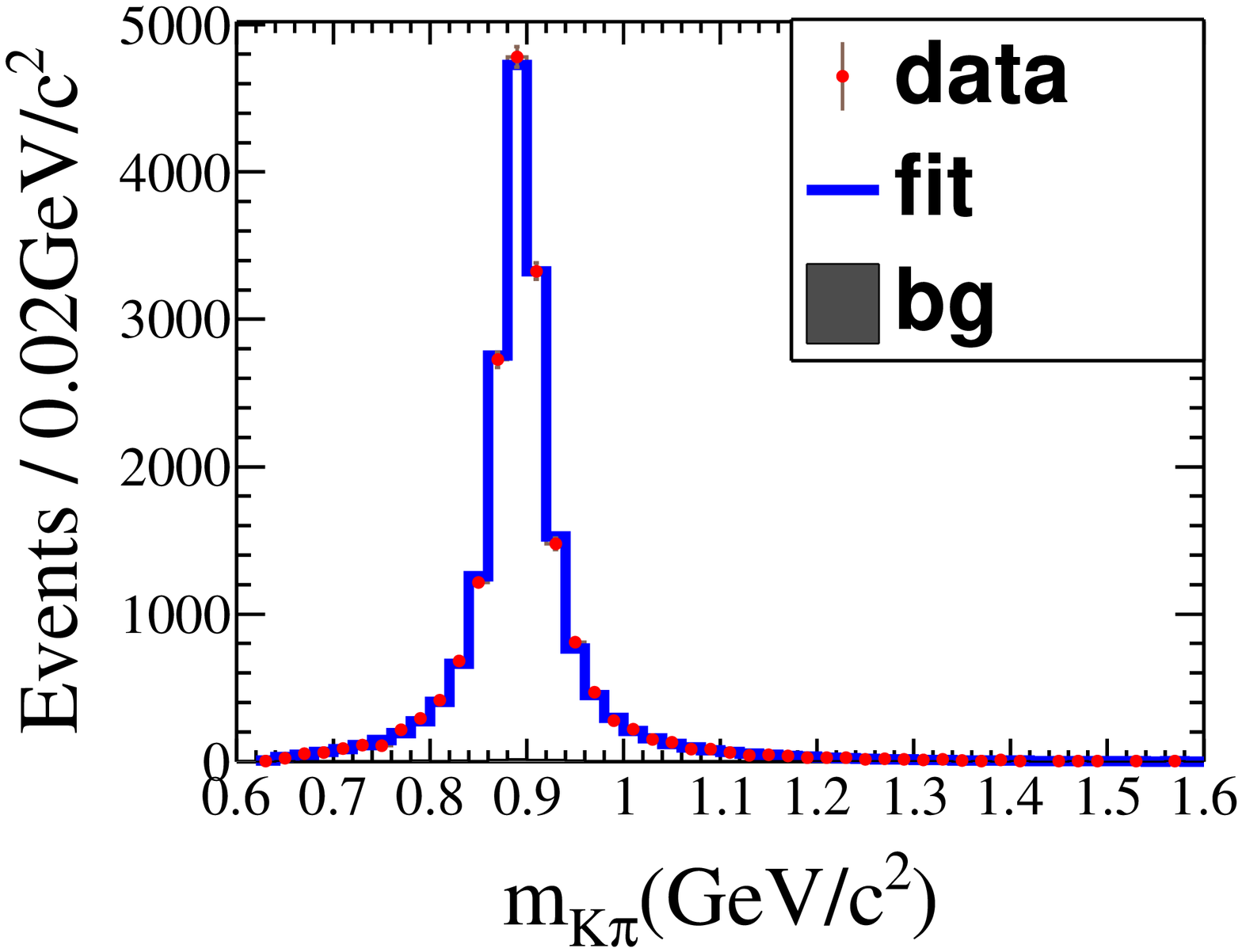}
\includegraphics[width=0.25\textwidth,height=0.18\textwidth]{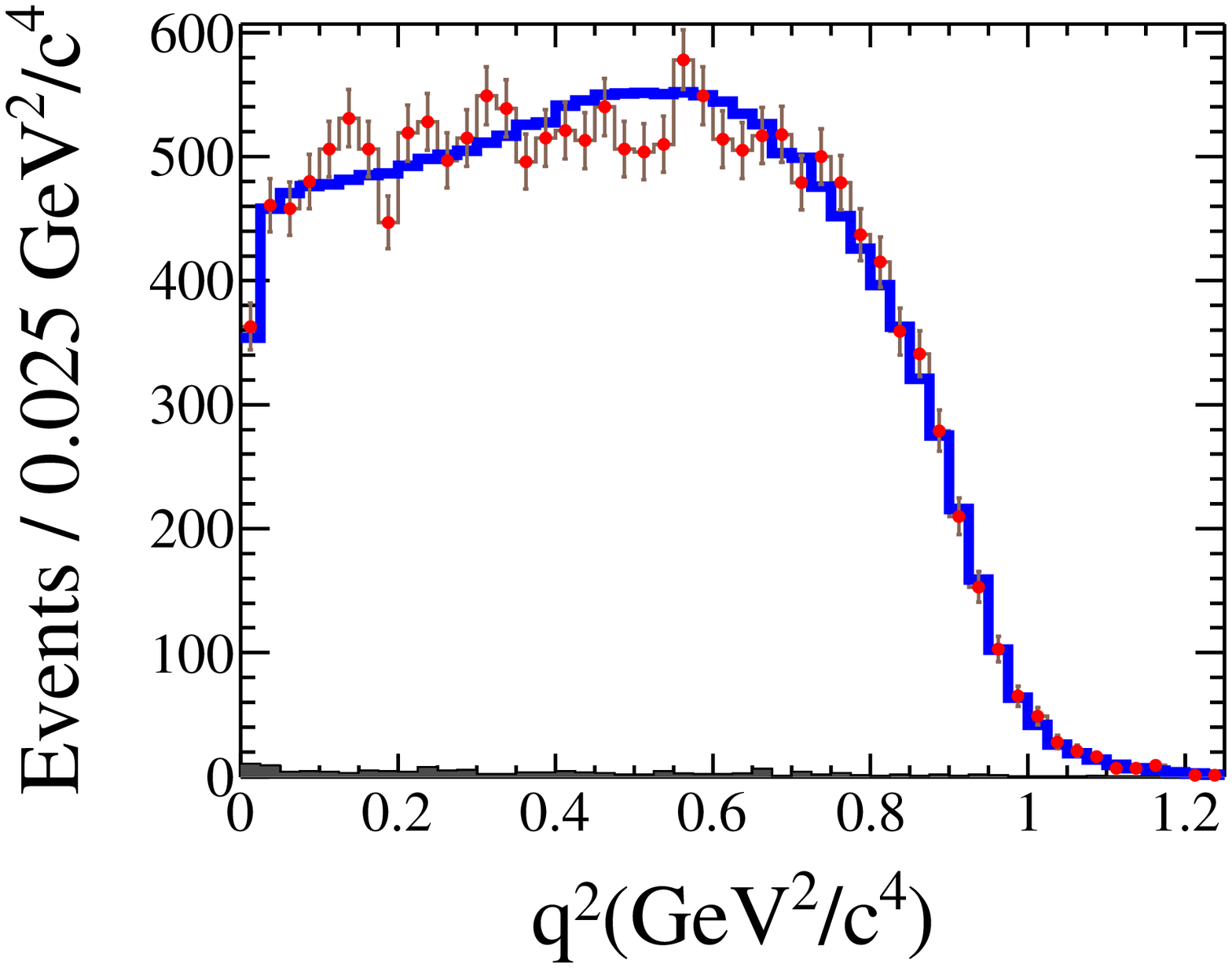}
\includegraphics[width=0.25\textwidth,height=0.18\textwidth]{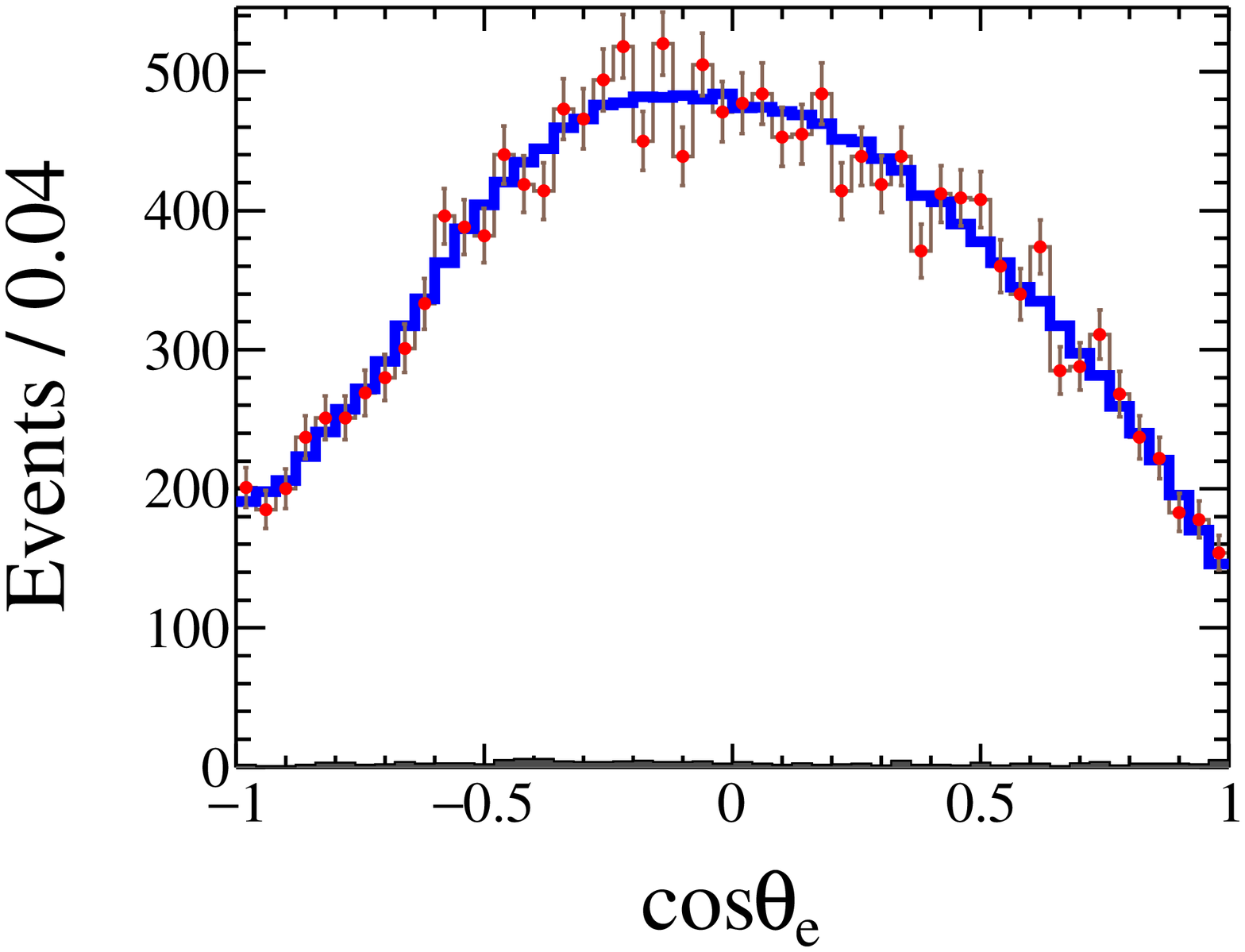}
\includegraphics[width=0.25\textwidth,height=0.18\textwidth]{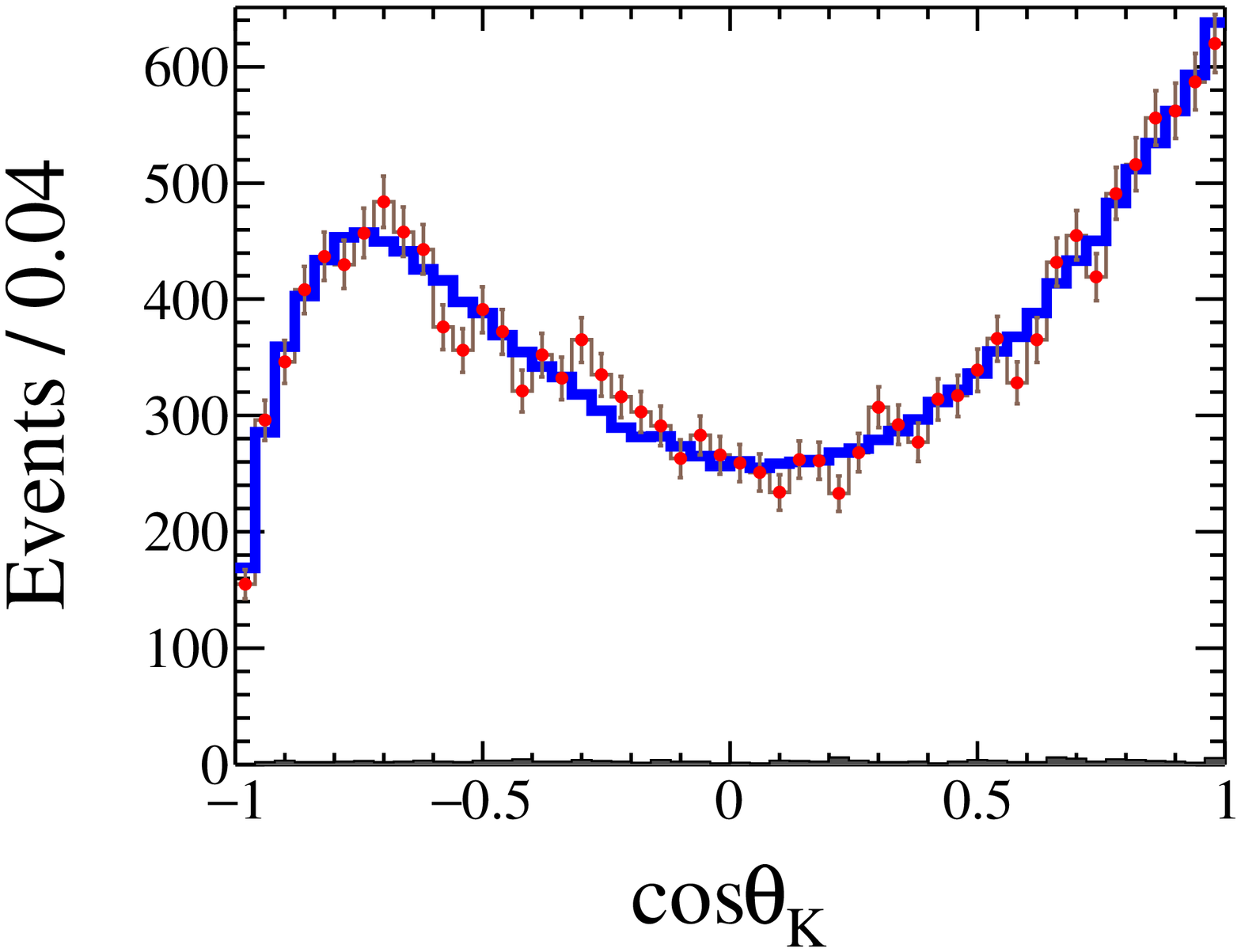}
\includegraphics[width=0.25\textwidth,height=0.18\textwidth]{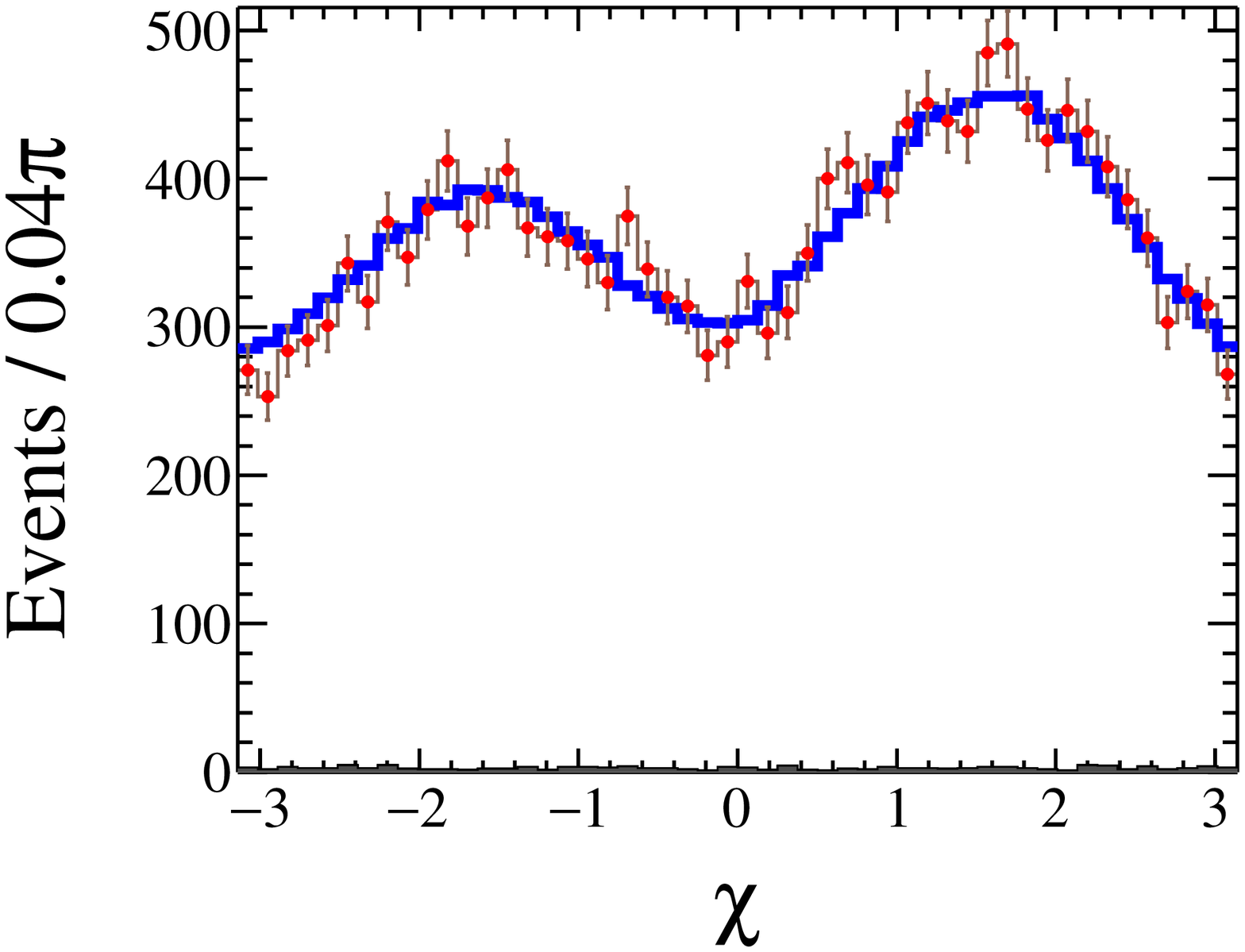}
\caption{
Projections of the kinematic variables of PWA for $D^+\to K^-\pi^+ e^+\nu_e$, where
$m_{K\pi}$ is the $K\pi$ mass,
$q^2$ is the $e\nu_e$ mass square,
$\theta_K$ is the angle between $\pi$ and $D$ momenta in the $K\pi$ rest frame,
$\theta_e$ is the angle between $\nu_e$ and $D$ momenta in the $e\nu_e$ rest frame and
$\chi$ is the angle between the two decay planes.
The dots with error bars are data, the blue curves are the weighted signal MC and
the hatched histograms are the simulated backgrounds.}
\label{fig:kstev}
\end{center}
\end{figure*}

Based on 18262 signals of $D^+\to K^-\pi^+e^+\nu_e$,
we determine the branching fraction
$$B(D^+\to K^-\pi^+e^+\nu_e)=(3.71\pm0.03\pm0.08)\%.$$
A partial wave analysis (PWA)
is performed on the selected candidates,
with results shown in Fig. \ref{fig:kstev}.
The PWA results show that the dominant $\bar K^{*0}$ component
is accompanied by an $S$-wave contribution accounting for
$(6.05\pm0.22\pm0.18)\%$ of the total rate, and other components can be
negligible.
We obtain the mass and width of $\bar K^{*0}(892)$
$M_{\bar K^{*0}(892)}=(894.60\pm0.25\pm0.08)$ MeV/$c^2$ and
$\Gamma_{\bar K^{*0}(892)}=(46.42\pm0.56\pm0.15)$ MeV/$c^2$,
the Blatt-Weisskopf parameter $r_{\rm BW}=3.07\pm0.26\pm0.11~({\rm GeV}/c)^{-1}$,
as well as the parameters of the hadronic form factors
$r_V=\frac{V(0)}{A_1(0)}=1.411\pm0.058\pm0.007$,
$r_2=\frac{A_2(0)}{A_1(0)}=0.788\pm0.042\pm0.008$,
$m_V=(1.81^{+0.25}_{-0.17}\pm0.02)$ MeV/$c^2$,
$m_A=(2.61^{+0.22}_{-0.17}\pm0.03)$ MeV/$c^2$,
$A_1(0)=0.585\pm0.011\pm0.017$.
Here, the first errors are statistical and the second systematic.

In the above PWA process,
the phase of the non-resonant background $\delta_S(m_{K\pi})$
is factorized by the LASS parameterizations,
and the helicity form factors $H_+(q^2,m_{K\pi})$, $H_-(q^2,m_{K\pi})$
and $H_0(q^2,m_{K\pi})$ are parameterized by the spectroscopic pole dominance (SPD) model.
We also make model-independent measurements
of the $\delta_S(m_{K\pi})$, and the helicity form factors, respectively.
The results are consistent with the expectations of
the corresponding models and previous measurements.

\subsubsection{$D^+\to \omega(\phi)e^+\nu_e$ \cite{bes3_omgev}}
\begin{figure*}[htbp]
\begin{center}
\includegraphics[width=0.75\textwidth,height=0.35\textwidth]{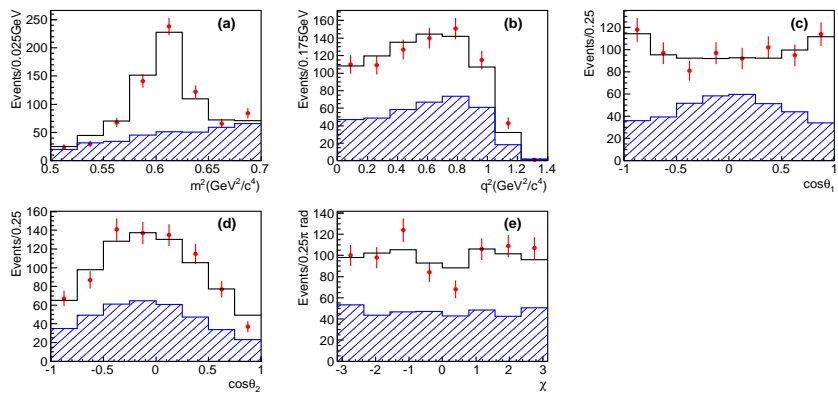}
\caption{
Projections of the kinematic variables of amplitude analysis for $D^+\to \omega e^+\nu_e$,
where the dots with error bars are data, the histograms are the fitted results and
the hatched histograms are the simulated backgrounds.}
\label{fig:omgev}
\end{center}
\end{figure*}

Based on $491\pm32$ signals of $D^+\to \omega e^+\nu_e$,
we determine the branching fraction
$$B(D^+\to \omega e^+\nu_e)=(1.63\pm0.11_{\rm stat.}\pm0.08_{\rm sys.})\times 10^{-3},$$
which is consistent with previous measurements but with better precision.
We perform amplitude analysis of the selected candidates,
with results shown in Fig. \ref{fig:omgev}.
We obtain the ratios of the hadronic form factors to be
$r_V=\frac{V(0)}{A_1(0)}=1.24\pm0.09_{\rm stat.}\pm0.06_{\rm sys.}$ and
$r_2=\frac{A_2(0)}{A_1(0)}=1.05\pm0.15_{\rm stat.}\pm0.05_{\rm sys.}$.

Also, we search for $D^+\to \phi e^+\nu_e$, but do not find obvious signal.
So, we set the upper limit on the branching fraction for $D^+\to \phi e^+\nu_e$
to be $1.3\times 10^{-5}$ at 90\% Confidence Level, which is significantly better than previous searches.

\section{Summary}

By analyses of the
leptonic decay $D^+ \to \mu^+\nu_\mu$ and the semileptonic
decays $D^0\to K(\pi)^-e^+\nu_e$,
$D^+\to K^0_L e^+\nu_e$,
$D^+\to K^-\pi^+e^+\nu_e$ and
$D^+\to \omega(\phi)e^+\nu_e$ from 2.92 fb$^{-1}$
data taken at $\sqrt s=$3.773 GeV with the BESIII detector,
we extract the $D^+$ decay constant,
the hadronic form factors and
the quark mixing matrix elements $|V_{cs(d)}|$.
These provide key experimental data to
validate the LQCD calculations of the $D^+$ decay constant and
the hadronic form factors
and to test the unitarity of the quark mixing matrix
at higher accuracies.

\Acknowledgements
I would like to thank for the support of
the National Natural Science Foundation of China (NSFC) under
Contracts No. 10935007 and No. 11305180, and the Ministry of Science
and Technology of China (973 by MOST) under Contracts
No. 2009CB825200 and No. 2015CB856700.

\setlength{\baselineskip}{12pt}

\end{document}